\documentclass[aps,prx,english,superscriptaddress,twocolumn,tightenlines,showpacs,longbibliography,floatfix,amsfonts]{revtex4-1}
\usepackage[T1]{fontenc}
\usepackage[utf8]{inputenc}
\usepackage[normalem]{ulem}
\usepackage{textcomp}
\usepackage{graphicx,float}
\usepackage[dvipsnames,svgnames]{xcolor}
\usepackage{mathrsfs,mathtools,xfrac}
\usepackage{amsmath,amssymb,amsthm,amscd,bm}
\usepackage{txfonts} 
\usepackage{multirow,comment}
\usepackage{array,booktabs,minibox}
\newcolumntype{P}[1]{>{\centering\arraybackslash}p{#1}} 
\usepackage[colorlinks,bookmarks=false,citecolor=blue,linkcolor=blue,urlcolor=blue,filecolor=blue]{hyperref}
\usepackage{wasysym}
\usepackage{makecell}

\newcolumntype{C}{>{$}c<{$}}
\AtBeginDocument{
\heavyrulewidth=.08em
\lightrulewidth=.05em
\cmidrulewidth=.03em
\belowrulesep=.65ex
\belowbottomsep=0pt
\aboverulesep=.4ex
\abovetopsep=0pt
\cmidrulesep=\doublerulesep
\cmidrulekern=.5em
\defaultaddspace=.5em
}

\newcommand{\eref}[1]{Eq.~\eqref{#1}}
\newcommand{\sref}[1]{Sec.~\ref{#1}}
\newcommand{\fref}[1]{Fig.~\ref{#1}}
\newcommand{\tref}[1]{Table.~\ref{#1}}
\newcommand{\aref}[1]{Appendix.~\ref{#1}}

\begin{document}

\title{Taming Spacetime Overhead and Design Complexity in Distributed Fault-Tolerant  \\ Superconducting Quantum Computation}

\author{Qinjing Yu}
\email{qinjing.yu@mail.ustc.edu.cn}
\affiliation{Hefei National Research Center for Physical Sciences at the Microscale and School of Physical Sciences, University of Science and Technology of China, Hefei 230026, China}
\affiliation{Hefei National Laboratory, University of Science and Technology of China, Hefei, 230088, China}
\affiliation{Shanghai Research Center for Quantum Science and CAS Center for Excellence in Quantum Information and Quantum Physics, University of Science and Technology of China, Shanghai 201315, China}

\author{Ke Liu}
\email{ke.liu@ustc.edu.cn}
\affiliation{Hefei National Research Center for Physical Sciences at the Microscale and School of Physical Sciences, University of Science and Technology of China, Hefei 230026, China}
\affiliation{Shanghai Research Center for Quantum Science and CAS Center for Excellence in Quantum Information and Quantum Physics, University of Science and Technology of China, Shanghai 201315, China}

\date{\today}

\begin{abstract}
Scaling fault-tolerant superconducting quantum computers will likely require distributed architectures built from multiple manufacturable quantum processing units. 
A central question is whether noisy and slow inter-chip operations impose substantial spacetime overhead or orchestration burdens compared with monolithic architectures.
To answer this question, we present a hardware-grounded architectural co-design together with a comprehensive resource-estimation protocol for surface-code-based modular processors.
The design confines inter-chip latency and noise to module boundaries, preventing slow, noisy links from inducing prohibitive spacetime overhead or becoming a global orchestration bottleneck.
The resource-estimation protocol integrates hardware constraints and circuit-level error-correction performance into  utility-scale algorithmic cost estimates, enabling a controlled assessment of different architectures.
Using RSA-2048 factorization as a demanding benchmark, we estimate resources under experimentally anchored parameters and realistic superconducting hardware constraints.
Compared with a large, ideal monolithic baseline, the resulting distributed architecture requires only modest additional resource overhead in both qubit count and execution time.
More importantly, the overhead is nearly scale-invariant across a broad module-capacity window, decoupling chip size from global performance. 
This decoupling turns module capacity from a finely tuned architectural parameter into a flexible engineering degree of freedom, allowing chip sizes to be set by manufacturability and control-packaging constraints rather than architectural fine-tuning.
These results establish a viable route to distributed fault-tolerant superconducting quantum computation that scales without prohibitive resource growth or heavy orchestration burden.
\end{abstract}

\maketitle

\begin{figure*}[htbp]
    \centering
    \resizebox{\linewidth}{!}{
        \includegraphics{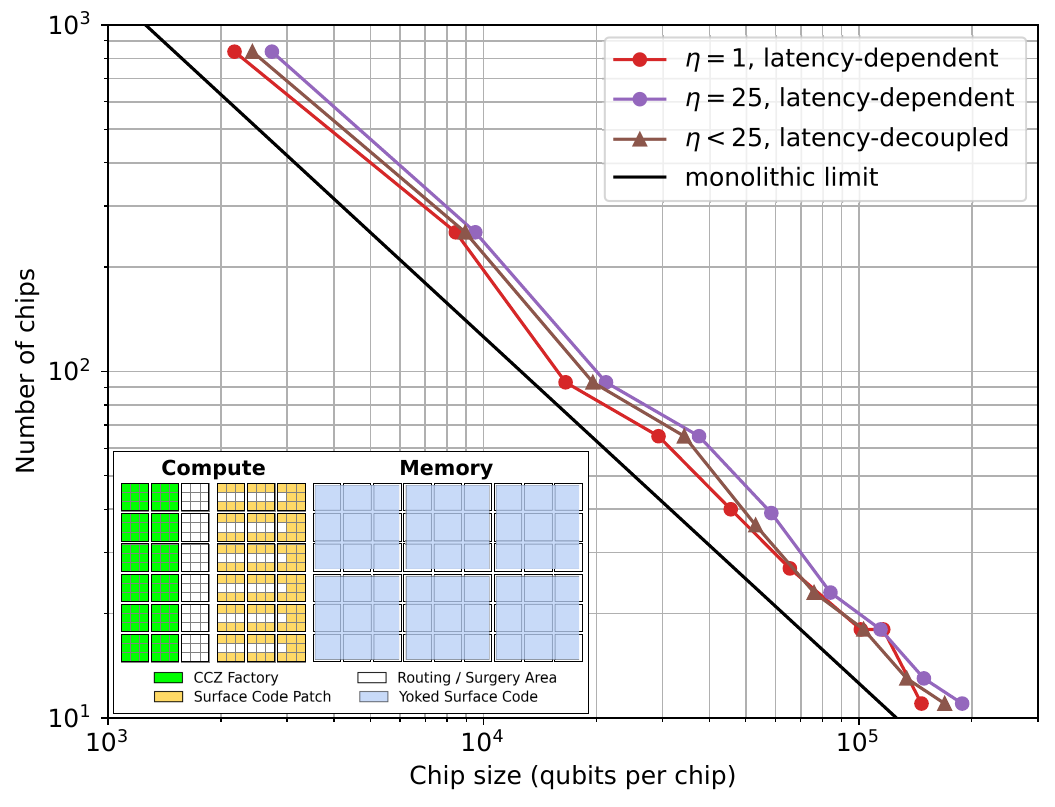}
        \hfill
        \includegraphics{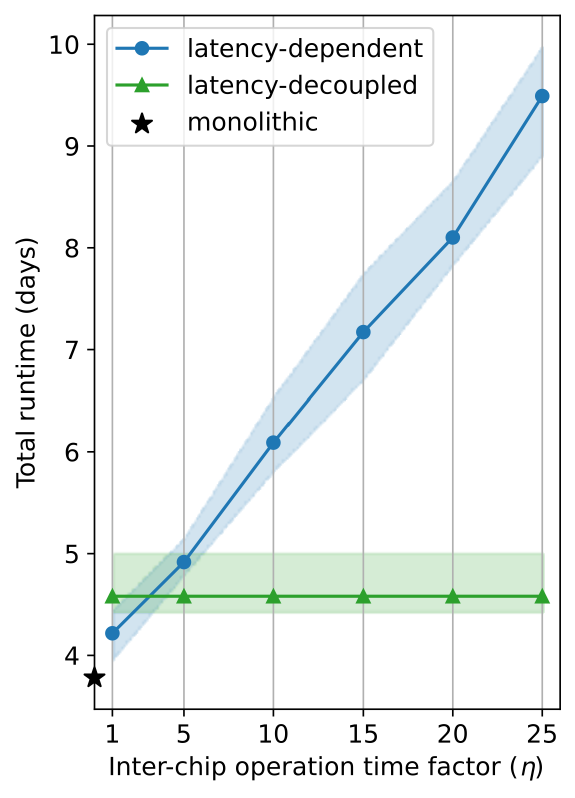}
    }
    \caption{Spacetime overhead of distributed architectures for factoring 2048-bit RSA integers. A hardware-motivated noise model is considered, with physical error rate $10^{-3}$ and $10^{-2}$ for intra-chip and inter-chip operations, respectively.
    Left: Required number of QPUs as a function of chip capacity (qubits per QPU). The solid black line indicates the monolithic limit and effectively represents a constant total qubit count. Colored lines correspond to distributed architectures with various chip capacities, for different inter-chip operation time factors ($\eta$) and interface circuit protocols. Variations in chip size have a minimal impact on the total qubit count in the distributed architecture. Inset: Schematic chip layout illustrating the functional partition of the distributed architecture into computation and memory zones. Right: total runtime of the algorithm as a function of the inter-chip operation time factor ($\eta$). The black star indicates the runtime of the monolithic architecture. The blue and green lines depict the runtime for the latency-dependent protocol and the latency-decoupled protocol, respectively. Shaded areas denote the upper and lower bounds across all searched configurations.
    }
    \label{fig:spacetime_overhead_1e-3}
\end{figure*}

\section{Introduction}\label{sec:introduction}

The pursuit of fault-tolerant quantum computation is progressively transitioning from proof-of-principle demonstrations to utility-scale realization. Recent experimental validations of below-threshold error correction (QEC) and universal logical operations have underpinned the subsequent phase of systemic scaling~\cite{google_quantum_ai_suppressing_2023,google_quantum_ai_and_collaborators_quantum_2025,postler_demonstration_2022,bluvstein_logical_2024}. However, executing algorithms of practical value, such as factoring RSA-2048 or simulating quantum chemistry, typically requires on the order of millions of physical qubits—a scale that is engineeringly not realistic for a single quantum processing unit (QPU)~\cite{gidney_how_2021,gidney_how_2025,su_fault-tolerant_2021,mohseniHowBuildQuantum2024,bravyiFutureQuantumComputing2022a}. Distributed fault-tolerant quantum computation (DFTQC) has emerged as a leading route to overcome the scaling bottleneck. The core strategy of DFTQC is to create a massive-scale quantum system by integrating multiple smaller manufacturable QPUs~\cite{nickerson_topological_2013,li_hierarchical_2016,singh_modular_2024,jacinto_network_2025,chandra_architectural_2025,naito_network-based_2026,haug_lattice_2025}. The distributed scheme effectively circumvents the fabrication hurdles, control and calibration complexity, and power capacity constraints inherent to large monolithic chips~\cite{mohseniHowBuildQuantum2024,bravyiFutureQuantumComputing2022a,gold_entanglement_2021,niu_low-loss_2023,heya_randomized_2025,song_realization_2024,qiu_deterministic_2025,franke_rents_2019}.

Growing research efforts are currently devoted to the development of distributed quantum computing architectures. On the theoretical front, recent analyses have systematically established the noise resilience of inter-chip interfaces, revealing an error threshold at the boundary nearly one order of magnitude higher than within the internal bulk~\cite{li_hierarchical_2016,ramette_fault-tolerant_2024,shalby_optimized_2025}. Moreover, various remote gate strategies and entanglement distribution protocols have been proposed across diverse hardware platforms, including superconducting circuits, trapped ions, neutral atoms, and solid-state qubits. At a macroscopic level, different modular topologies have been proposed to optimize the organization of networked QPUs, ranging from static patch-based structures to dynamically switched networks~\cite{chandra_architectural_2025,filippov_architecting_2025,naito_network-based_2026}.
On the experimental side, the foundational hardware infrastructure required to support distributed quantum computing has been implemented and is undergoing rapid refinement~\cite{magnard_microwave_2020,storz_loophole-free_2023,conner_multi-chip_2023,heya_multi-module_2025,krutyanskiy_entanglement_2023,main_distributed_2025,pompili_multinode_2021,stolk_metropolitan_2024,knaut_entanglement_2024,warner_coherent_2025}.
Driven by these advancements, industry roadmaps are increasingly aligning with DFTQC schemes~\cite{mohseniHowBuildQuantum2024,bravyiFutureQuantumComputing2022a}.

While these studies establish the feasibility of key DFTQC primitives, a central concern remains: whether distributed modular architecture would introduce prohibitive spacetime overhead, or whether maintaining near-monolithic performance would require substantial architectural complexity or heavy system-level orchestration. Specifically, if the additional qubit count and execution time introduced by noisy and slow inter-chip operations vastly exceed the baseline cost of monolithic systems, the intended scalability advantage can be severely compromised. Moreover, since standard QEC strategies have been developed primarily for monolithic processors, their direct application to distributed architectures may overlook realistic hardware constraints and lead to critical performance bottlenecks. As the existing literature is often fragmented between microscopic interconnect studies and macroscopic topology design, there remains a clear gap for a system-level investigation that synthesizes physical hardware constraints, fault-tolerant QEC protocols, and large-scale algorithm execution.

In this Article, we present an architectural co-design for distributed superconducting quantum computation and conduct a comprehensive resource estimation for factoring RSA-2048. Such analyses are sensitive to the underlying hardware parameters and physical assumptions. Thus, we focus on superconducting chips with nearest-neighbor connectivity and surface-code-based schemes. 
Although quantum low-density parity-check (qLDPC) based schemes may in principle substantially reduce qubit overheads, their practical advantage remains sensitive to open assumptions about noise models, logical operations, decoding efficiency, intrinsic code scalability, and the fabrication and control complexity of nonlocal connectivity.
These factors can significantly affect the performance gains predicted by idealized qLDPC resource estimates in realistic implementations.
To keep the estimate controlled and hardware-relevant, we work in an experimentally anchored regime directly tied to current superconducting hardware platforms \cite{webster_pinnacle_2026,yoderTourGrossModular2025, bravyiHighthresholdLowoverheadFaulttolerant2024}. 
The need for such an analysis is pressing as superconducting quantum processors reach the hundred-qubit scale, turning the transition from isolated devices to scalable fault-tolerant modular systems into an increasingly important question.

We demonstrate that architectural co-design can mitigate the fault-tolerance overheads induced by modularity, yielding resource costs comparable to those of a large, idealized monolithic processor.
Under practical physical parameters---an intra-chip two-qubit gate error rate $p=10^{-3}$, inter-chip operations that are $10\times$ noisier, and inter-chip gates that are up to $25\times$ slower---DFTQC requires about $2.0$ million physical qubits and an expected runtime of $4.4$ days for factoring RSA-2048. By comparison, a monolithic architecture using the same QEC strategies requires about $1.3$ million physical qubits and $3.4$ days~\footnote{These monolithic comparison values differ from the rounded figures in Ref.~\cite{gidney_how_2025}: the larger qubit count comes from using the SI1000 noise model, which raises the monolithic distance to 27, and the shorter runtime comes from not applying rounded-up subroutine times~\cite{gidney_answer_2025}; see \tref{tab:rsa2048_subroutine_runtimes} and \tref{tab:rsa2048_resource_estimates}.}. 
In terms of QEC requirements, this corresponds to only a modest increase in the surface-code distance, from $d=27$ to $d=31$.
Thus, distributed execution introduces only about $60\%$ additional physical-qubit overhead and $30\%$ additional runtime. These estimates are made on a conservative basis, as individual QPU layouts reserve substantial physical-qubit slack. If chip sizes are tailored to the actually used qubits, the required qubit count can be reduced to about $1.5$ million, only about $30\%$ above the monolithic baseline.

More importantly, the additional overhead is nearly {\it scale-invariant} over a broad range of module capacities.
This behavior is illustrated in \fref{fig:spacetime_overhead_1e-3} over an experimentally relevant capacity window, ranging from a few thousand to nearly two hundred thousand physical qubits per chip.
The underlying reason is that our design keeps modularity overhead as a local boundary cost: inter-chip latency and noise are confined to QPU boundaries, preventing slow and noisy links from becoming a global orchestration bottleneck. 
The same trend persists in the lower-error regime $p=10^{-4}$, indicating that the weak dependence of spacetime overhead on module size is not tied to a particular error point.
Moreover, this near scale-invariant behavior should extend to other computational tasks implemented with surface-code-based error correction.
Indeed, after compilation into a universal logical gate set, cross-module operations reduce to multi-body Pauli-measurement primitives, whose implementation is algorithm-agnostic.

The practical significance of this finding transcends the numerical values themselves and carries multiple implications. At the hardware level, this flexibility makes distributed fault tolerance compatible with manufacturability, since module capacities can be chosen to balance fabrication yield, testing complexity, and device-level reliability. At the system-integration level, it allows chip size to be adapted to wiring, calibration, packaging, and cryogenic-control constraints rather than being dictated by a sharp fault-tolerance penalty. At the architectural level, the result provides design freedom, as chip size can remain a flexible architectural choice rather than being fine-tuned.

Our work substantially mitigates the risk of prohibitive resource scaling and largely removes the performance gap due to modularity. The weak dependence of global overhead on chip size greatly simplifies system-level optimization, avoiding extensive chip-size-specific fine-tuning of architectural orchestration and network organization. We use RSA-2048 factorization as a concrete and demanding benchmark, but our resource-estimation protocol is readily extensible to other utility-scale algorithms. 
The resulting estimates are tied to realistic nearest-neighbor superconducting hardware constraints and are grounded in circuit-level simulations and experiment-motivated noise models.
Moreover, thousand-qubit-scale modules align with near-term superconducting roadmaps while leveraging mature industrial 6-inch and 12-inch wafer fabrication.

The remainder of this Article is organized as follows. \sref{sec:architecture_design} summarizes the key physical and architectural factors that control distributed fault-tolerant overhead. \sref{sec:resource_estimation} formulates a general resource-estimation procedure for DFTQC. \sref{sec:circuit_and_error_modeling} develops the latency-decoupled inter-chip syndrome-extraction protocol and the corresponding logical-error model. \sref{sec:factorization_resource_estimation} applies the full procedure to RSA-2048 factorization in both a near-term physical error regime and a lower physical error regime, showing that the extra distributed overhead remains modest and nearly scale-invariant across a broad range of chip capacities. We conclude in \sref{sec:conclusion} with an outlook. Technical details on the noise model, fitting methodology, RSA integer factorization subroutines, runtime analysis, and layouts of yoked surface codes are provided in Appendices.

\section{Key factors affecting distributed architecture design}\label{sec:architecture_design}

A FTQC architecture typically organizes a full system into dedicated functional zones including computation, memory, and classical control~\cite{gidney_how_2025, webster_pinnacle_2026,mundadaHeterogeneousArchitecturesEnable2026}.

In a DFTQC architecture, these functions are distributed across physically separate QPUs of limited size, interconnected through quantum interconnects. Computation modules execute both Clifford and non-Clifford fault tolerant logical operations, memory modules store idle logical qubits under active error correction, and classical control orchestrates scheduling, decoding, and data routing across the entire system. The quantum interconnects between QPUs---realized, for example, through shared Bell pairs consumed by teleportation-based protocols---make system-wide quantum information processing possible by mediating the nonlocal stabilizer measurements, distributed lattice surgery, or logical state teleportation that stitch individually small modules into a coherent, large-scale fault-tolerant processor.

Because DFTQC architectures rely fundamentally on inter-chip quantum links, the properties of these interfaces largely dictate the additional overhead incurred by modularity. Although the physical realization of quantum interconnects varies across hardware platforms, their impact on system performance is fundamentally governed by three critical factors: inter-chip operation fidelity, communication latency, and the layout of logical qubits.

The fidelity of inter-chip operations is arguably the most obvious bottleneck. Owing to the additional loss and decoherence intrinsic to remote quantum interconnects, inter-chip entangling operations generally exhibit higher error rates than local bulk operations, whether they are implemented through direct remote entangling gates or through other mediated entanglement-generation protocols. These elevated error rates can be quantified by an interface-to-bulk error ratio $\gamma$. A large $\gamma$ can severely degrade the fidelity of logical qubits and gates, thereby imposing a substantial space-time overhead on the overall computation. To counteract this effect, modular architectures must employ specialized boundary gadgets that preserve the performance of the underlying quantum error-correcting code.

Inter-chip operations also take longer than local gates. We characterize this difference by a time factor $\eta$, defined as the ratio of inter-chip to intra-chip operation duration. A large $\eta$ increases the effective QEC cycle time and slows both Clifford and non-Clifford logical operations. This not only increases the temporal overhead, but also leads to longer execution time, which in turn accumulates more idle errors and the spatial overhead. Moreover, the difference in QEC cycle time at the interface can desynchronize them from the rest of the system, complicating the scheduling strategies to maintain logical synchronization~\cite{maurya_synchronization_2025}. Mitigating additional spacetime overhead caused by communication latency therefore requires minimizing the number of inter-chip operations and maximizing concurrency between inter-chip and intra-chip operations.

\begin{figure*}[!t]
    \centering
    \includegraphics[width=0.75\linewidth]{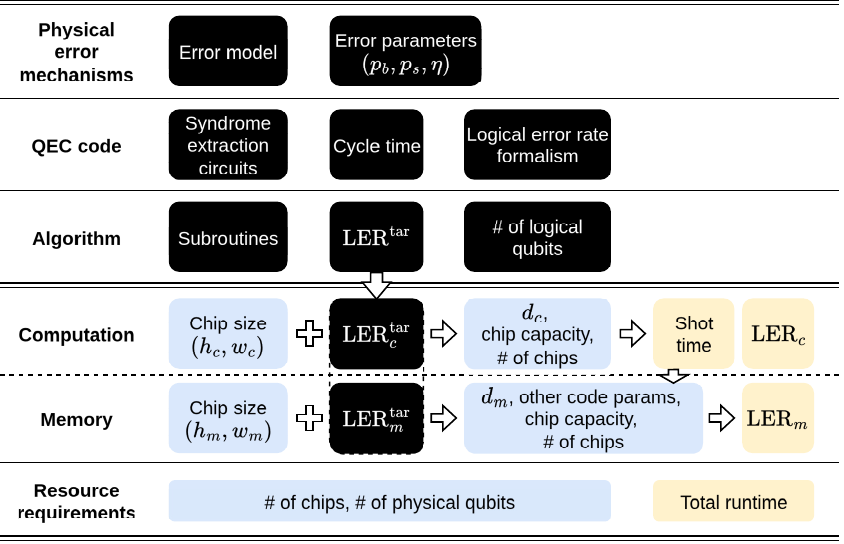}
    \caption{General procedure for resource estimation in the DFTQC architecture. The parameters highlighted in black are prerequisite inputs, while those highlighted in blue and yellow represent space cost and time cost, respectively.}\label{fig:procedure}
\end{figure*}

Furthermore, from a macroscopic viewpoint, modularity can potentially affect the optimization of logical qubit layout and fault tolerant functional zones, thereby introducing additional design complexity. The system performance can depend on both chip size and topology of chip network~\cite{naito_network-based_2026}. Meanwhile, different fault tolerant gadgets can be affected differently by inter-chip operations. For instance, magic state production is extremely sensitive to the elevated error rates of inter-chip operations because the success rate decreases exponentially with error rate~\cite{gidney_magic_2024}. These considerations therefore require co-optimizing logical layout and architectural partitioning to localize communication-intensive operations and avoid unnecessary inter-chip traffic.

The primary objective of DFTQC architecture design is to minimize the additional space-time overhead introduced by modularity and inter-chip quantum interconnects. As we shall see, this challenge can be largely resolved at the QEC level. In particular, appropriately designed boundary gadgets can absorb much of the complexity that would otherwise have to be managed through system-level optimization, bringing both the overall design complexity and resource overhead close to those of a monolithic architecture.

\section{Procedure of resource estimation}\label{sec:resource_estimation}

Resource requirements for FTQC in monolithic architectures has been extensively studied, especially on the superconducting platform, where all computation and storage are supported on a single monolithic chip with a contiguous array of physical qubits ~\cite{gidney_how_2021, gidney_how_2025,hugginsFLuidAllocationSurface2025}. 

In a distributed architecture, the partitioning of logical qubits across physically separate QPUs introduces more factors as discussed in \sref{sec:architecture_design}, and the resource estimation must be extended to account for these factors so as to accurately capture the overhead of modularity. The interface-specific noise channels must be tracked separately and fed into the logical error rate formalism alongside the bulk error. The longer duration of inter-chip operations requires explicitly accounting when evaluating the logical cycle time. The finite chip size imposes a geometric constraint that couples the code distance, the number of inter-chip interfaces, and the logical qubit layout. It determines how many logical operations each interface involves, and consequently how much of the error budget is consumed by inter-chip contributions. As a result, the chip size becomes an optimization variable that trades off chip capacity against interface overhead---a degree of freedom with no counterpart in the monolithic case. Together, these extensions make the resource estimation for DFTQC a more complex problem that requires a more comprehensive procedure to synthesize the various factors and determine the optimal design parameters.

Without loss of generality, we assume that each QPU contains intact logical qubits, and inter-chip coupling occurs only through lattice surgery between logical patches. This is a natural assumption for superconducting platforms, where chips with capacities on the order of $10^3$ physical qubits---sufficient for the footprint of a high-fidelity logical qubit---will be available in the near future, while much smaller chips would aggravate wiring and control complexity. A collateral benefit is that magic-state injection or cultivation is performed locally within each QPU, which is critically important because magic-state preparation is especially sensitive to elevated inter-chip error rates due to its exponentially decreasing success probability with physical noise~\cite{liMagicStatesFidelity2015, gidney_magic_2024}.

Among the functional zones described in \sref{sec:architecture_design}, computation and memory are the most critical for resource estimation: the computation zone determines the cost of executing algorithmic subroutines, including Clifford operations and the generation of magic states, while the memory zone determines the cost of maintaining idle logical qubits for the duration set by the computation. The overhead of quantum interconnects and classical control can be accounted for within these two zones. Accordingly, our procedure is organized around these two zones.

Moreover, the procedure is generic and not restricted to a specific algorithm or QEC code. The reason is that any quantum circuit can be compiled into the Clifford+T gate set; Clifford gates are realized by multi-body logical Pauli measurements via lattice surgery, while each T gate is implemented by consuming a magic state through a Pauli measurement. Magic states themselves are produced by non-fault-tolerant injection followed by distillation, which is again carried out via Pauli measurements. Consequently, the entire computation reduces to multi-body logical Pauli measurements and non-fault-tolerant magic-state preparations~\cite{litinskiGameSurfaceCodes2019}. In a DFTQC architecture, inter-chip couplings arise only in the lattice surgery used to implement the former, while magic-state preparation remains local to each QPU. Therefore, algorithm dependence enters mainly through the counts and layout of these operations, rather than through additional algorithm-specific communication strategies. Likewise, the procedure is modular with respect to QEC codes: the computation and memory zones can independently adopt different chip sizes, code families, and optimization criteria.

The resulting procedure is illustrated in \fref{fig:procedure} and proceeds through three layers.

\paragraph{Prerequisite inputs (black in \fref{fig:procedure}).}
Three categories of prerequisite inputs are required:
(i)~\emph{Physical error mechanisms}, which specify the error model (e.g., depolarizing, biased) and its parameters---in our case, the bulk error rate $p_b$, the interface error rate $p_s$, and the inter-chip time factor $\eta$;
(ii)~\emph{QEC code specifications}, which define the syndrome extraction circuits, the QEC cycle time, and the logical error rate formalism (e.g., the ansatz of \eref{eq:ler_ansatz});
(iii)~\emph{Algorithm specifications}, which provide the set of subroutines composing the target algorithm, the target logical error rate $\text{LER}^{\text{tar}}$ for the overall computation, and the number of logical qubits.

\paragraph{Chip-level partitioning.}
This layer optimizes the code parameters, chip geometry, number of chips, and the resulting logical error rates for both the computation and memory zones through grid scanning. Each zone is parameterized by its own chip geometry and target logical error rate.

For the \emph{computation} zone, the inputs are the chip geometry $(h_c, w_c)$---the number of logical qubits along the height and width of a computation chip---and the target logical error rate $\text{LER}^{\text{tar}}_c$. Given these inputs, the procedure determines the minimum code distance $d_c$ that satisfies $\text{LER}^{\text{tar}}_c$ by evaluating the logical error rate formalism with both the bulk and inter-chip error contributions set by the chip geometry and the interface protocol. The chip capacity in physical qubits and the number of chips required are then derived from $(h_c, w_c, d_c)$ and the number of required logical qubits. The shot time is estimated by summing the duration of each subroutine weighted by its use count, while the logical error rate $\text{LER}_c$ is obtained by aggregating the error contribution of each subroutine. The shot time, highlighted in yellow in \fref{fig:procedure}, feeds into the memory zone.

For the \emph{memory} zone, the inputs are the chip geometry $(h_m, w_m)$ and the target logical error rate $\text{LER}^{\text{tar}}_m$. Crucially, the parameters in the memory zone depends on the computation zone because the required storage duration is set by the shot time. The procedure determines the code distance $d_m$ and, if a different QEC code is used, additional code parameters such as the outer-code dimension for yoked surface codes \cite{gidney_yoked_2025}. The memory logical error rate $\text{LER}_m$ is then computed, taking into account the inter-chip interface contributions.

\paragraph{Aggregation of resource requirements.}
Finally, the total resource requirements are obtained by aggregating the space and time metrics from both zones. The space cost includes the total number of chips and the total number of physical qubits across computation and memory. The time cost is the total runtime, obtained from the shot time times the number of shots required for success divided by the total logical error rate composed of $\text{LER}_c$ and $\text{LER}_m$.

\begin{figure}[!t]
    \centering
    \includegraphics[width=\linewidth]{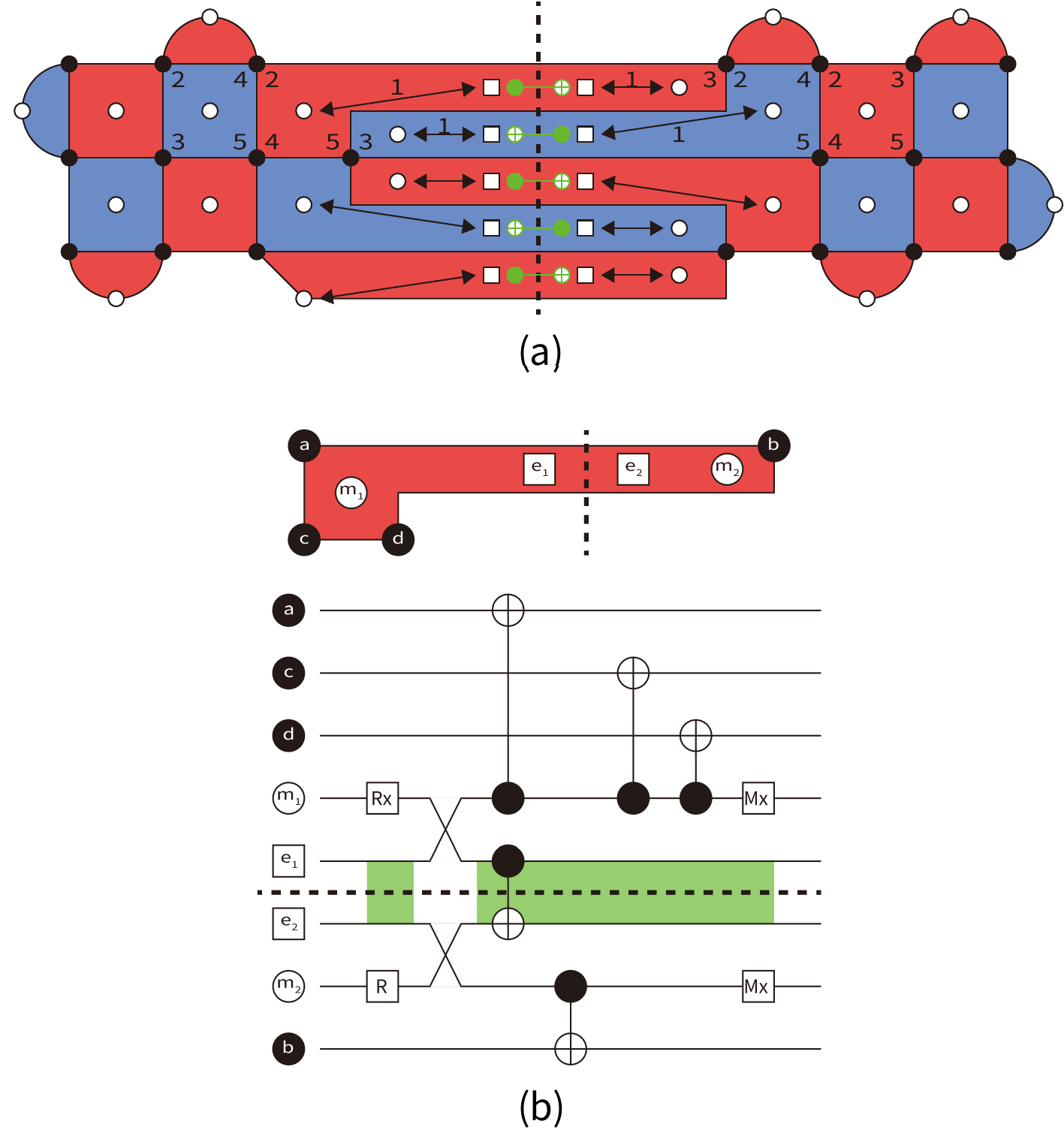}
    \caption{Physical qubit layouts at the interface and the corresponding gate sequences for inter-chip syndrome extraction. Blue (red) plaquettes denote Z(X)-stabilizers. The dashed line marks the interface between neighboring chips. The numbers indicate the order of two-qubit gates in one round of syndrome extraction. Inter-chip operations are shown in green. (a) Layout of our proposed circuit with buffer qubits (squares), which decouple intra-chip and inter-chip operation speeds while preserving code distance. Lines with arrows denote SWAP gates. (b) X-stabilizer measurement circuit with buffer qubits. The green region indicates the maximum allowed duration of the inter-chip operation in this protocol; it spans the entire QEC cycle except for the SWAP gate. Note that gate sizes are not proportional to their durations.}
    \label{fig:interchip_circuit}
\end{figure}

\section{Latency-decoupled Inter-chip \\ syndrome extraction}\label{sec:circuit_and_error_modeling}

In this section, we present a syndrome-extraction protocol that decouples intra-chip QEC cycles from slow inter-chip operations.

In state-of-the-art experiments, inter-chip two-qubit gate times range from approximately 100 ns~\cite{gold_entanglement_2021} to $1~\mu\text{s}$~\cite{heya_randomized_2025}, with best-case error rates around 0.9\%. In contrast, intra-chip two-qubit gates are typically much faster ($\sim$40 ns) and have lower error rates ($p_b \sim$0.1\%)~\cite{google_quantum_ai_and_collaborators_quantum_2025}. To connect the time cost of inter-chip operations to the QEC cycle time, which we take to be $1~\mu\text{s}$, we set the duration of an intra-chip two-qubit gate (CX or CZ) to 40 ns and write the inter-chip operation time as $\eta \times 40$ ns.
It is hence reasonable to assume $\eta$ ranging from 1 to 25. Slower inter-chip operations increase the cycle time of any logical patch involving an interface, with the precise increase determined by the interface circuit. For physical errors, we denote the depolarization probabilities of inter-chip and intra-chip two-qubit gates by $p_s$ and $p_b$, respectively, and assume that $p_s/p_b=\gamma=10$.

Throughout this section, we use the surface code at chip boundaries because of its high threshold and natural compatibility with superconducting hardware. The syndrome-extraction protocol is shown in \fref{fig:interchip_circuit}, which integrates a code-distance-preserving seam with a dedicated buffer layer. The seam adopts a hook-error-safe geometry with $2d-1$ interface ancilla qubits and unevenly split data qubits across the two joined boundaries~\cite{shalby_optimized_2025,jacinto_network_2025}. The buffer layer contains $4d-2$ ancillary qubits that generate Bell pairs concurrently with the intra-chip QEC cycle and then transfer them to neighboring measurement qubits. This protocol confines slow inter-chip operations to the buffer qubits and effectively decouples the intra-chip and inter-chip operation speeds, and is referred to as the \emph{latency-decoupled} protocol, in contrast to the conventional \emph{latency-dependent} protocol. In the worst case, we conservatively model the additional SWAP layer as three CX gates, increasing the cycle time from $1000$ ns to $(1000+3\times40)=1120$ ns, which is still far less than adding the extra latency of inter-chip operations directly to the cycle time. 

This latency mitigation is local to the boundary circuit: the design uses dedicated interface qubits, Bell-pair generation, and one SWAP layer, and shields the bulk stabilizer measurements. This locality is useful experimentally because the added hardware remains confined to the chip boundary and the timing requirement reduces to completing remote entanglement generation within one QEC cycle.

The resulting $\eta$-independent interface cycle time prevents the interface patches from forming a slow clock domain relative to the bulk. Because lattice surgery requires synchronized syndrome extraction cycles, avoiding this desynchronization also avoids additional synchronization-related machinery, including schedule barriers, runtime slack handling, and patch-phase bookkeeping in the control and compilation stack~\cite{maurya_synchronization_2025,mundadaHeterogeneousArchitecturesEnable2026}. Meanwhile, the protocol preserves the surface-code stabilizer structure, so it does not increase decoding complexity; only the interface noise model in the error analysis below is modified. At the architectural level, this converts a small local space cost into a system-level benefit: by insulating the bulk from slow inter-chip operations, it greatly mitigates the time cost of the distributed architecture and can even reduce the overall space overhead for large $\eta$.

To evaluate its performance, we benchmark a memory experiment on a $d\times 2d$ surface-code patch with an interface in the middle. Without loss of generality, we align the interface with the logical $X$ operators, which mainly increases the logical $X$ error rate, and therefore focus on memory of the logical $|0\rangle$ state. To model realistic noise in modular superconducting systems, we modify the SI1000 noise model~\cite{gidney_benchmarking_2022} to account for long-range connections. We distinguish between the bulk error rate $p_b$, which governs local operations, and the interface error rate $p_s$, which governs inter-chip operations. Meanwhile, the SWAP layer adds additional idle errors of $\eta \cdot p_b/10=3/10 p_b$ to the nonparticipating bulk qubits. The detailed noise model is described in \aref{app:noise_model}.

\begin{figure}[!t]
    \centering
    \includegraphics[width=\linewidth]{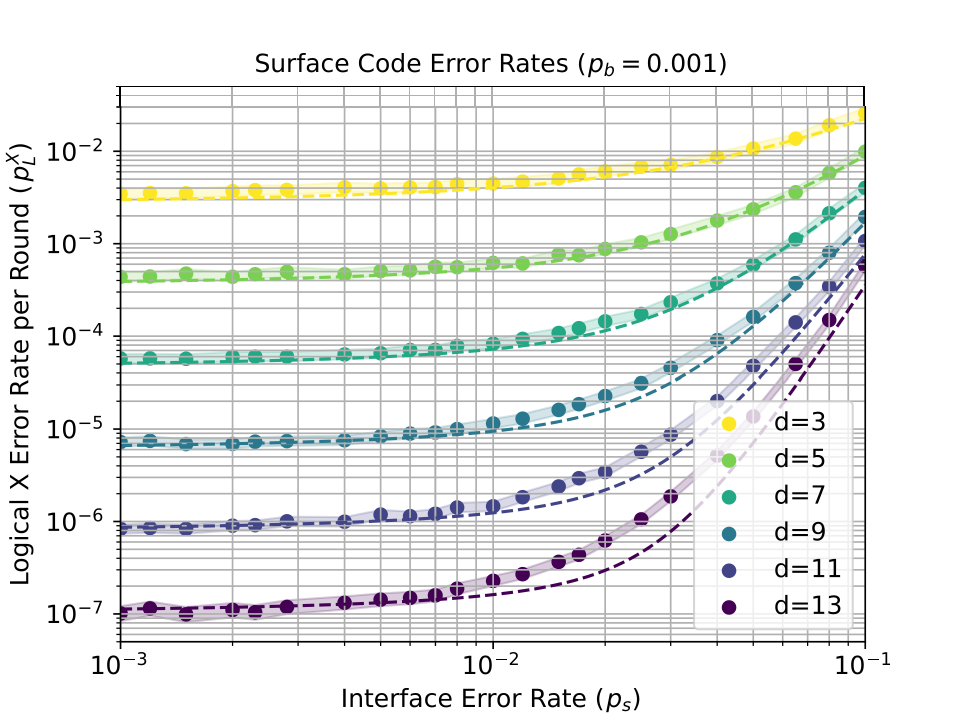}
    \caption{Logical X error rate per round $p_L^X$ as a function of the inter-chip physical error rate for various code distances under $\eta=3$ for the \emph{latency-decoupled} protocol. The bulk physical error rate is fixed at $p_b = 10^{-3}$. Dashed curves show fits to the ansatz of \eref{eq:ler_ansatz}. Shading represents hypothetical logical error rates with a Bayes factor of at most 1000 relative to the maximum-likelihood hypothesis probability, assuming a binomial distribution. For $p_s>10p_b=10^{-2}$, the logical error rates deviate slightly from the fitting curves, but the main resource estimation results are unaffected because we focus on the case of $p_s = 10p_b$.}
    \label{fig:buffer_fitting_results}
\end{figure}

To determine the code distance required to achieve a target logical error rate, we use the ansatz below~\cite{ramette_fault-tolerant_2024,jacinto_network_2025} to extrapolate the logical error rate per round $p_L^X$:
\begin{equation}
    p_L^X \approx \alpha_1\left(\frac{p_{s}}{p_{s}^*}\right)^{\frac{d+1}{2}} +\alpha_2\left(\frac{p_b}{p_b^*}\right)^{\frac{d+1}{2}}
    +\alpha_3 \sum_{1 \leq i \leq d}\left(\frac{p_{s}}{p_{s}^{**}}\right)^{\frac{i}{2}}\left(\frac{p_b}{p_b^*}\right)^{\frac{d+1-i}{2}}, \label{eq:ler_ansatz}
\end{equation}
where $p_s^*$ and $p_b^*$ are the thresholds associated with the interface and bulk error rates, and $p_s^{**}=p_s^*/[1+\alpha_c/(1-\sqrt{p_b/p_b^*})]$ is a pseudo-threshold for error chains containing both interface and bulk contributions. The quantities $\alpha_1$, $\alpha_2$, $\alpha_3$, and $\alpha_c$ are fitting constants. The ansatz separates the logical error rate into three distinct contributions: the first term represents interface-only errors, the second represents bulk-only errors, and the third captures mixed interface-bulk errors. By fitting the simulation data to this ansatz, we extract threshold values and coefficients for syndrome extractions with and without latency insulation. 
The results for the latency-decoupled protocol are shown in \fref{fig:buffer_fitting_results}, and more details of the fitting against various inter-chip time factors $\eta$ are provided in~\aref{app:fitting_details}.

The latency-decoupled protocol readily extends to more sophisticated surface-code-based encoding schemes.
As a representative example, we can analyze the performance of yoked surface codes~\cite{gidney_yoked_2025} in distributed architecture.
This scheme concatenates standard surface codes with outer high-density parity-check codes, thereby increasing the logical-qubit density in the memory zone. In particular, the 1D and square 2D outer codes belong to families of $[[n, n - 2, 2]]$ and $[[n^2, n^2 - 4n + 2, 4]]$ codes respectively.
We estimate the performance of different yoked-code schemes using the ansatz in \eref{eq:ler_ansatz}, with interface errors treated as a distinct additive channel. The logical error rates are approximated as:
\begin{align}
    P_{L,0}&\approx n_i \cdot p_{c,0}, \nonumber\\
    P_{L,1}&\approx n_i^2 \cdot r_o \cdot p_{c,1}^2, \nonumber\\
    P_{L,2}&\approx n_i^2 \cdot r_o \cdot p_{c,2}^4, \label{eq:yoke_ler}
\end{align}
where the subscripts $D=0,1,2$ denote the dimension of the outer code, $r_o$ is the number of outer-code rounds, $n_i$ is the number of inner-code patches, and $p_{c,k}$ is the logical error rate of a single inner-code patch during one outer-code round, obtained from simulation. When interface contributions are included, we add the interface error terms to $p_{c,k}$. Suppose there are $f$ interfaces involved in $r_l$ rounds of inter-chip syndrome extraction. The modified $p_{c,k}$ then becomes:
\begin{equation}
p_{c,k}' = p_{c,k} + \frac{r_l f}{n_i}p_L^s,\label{eq:dist_yoke_single_patch_ler}
\end{equation}
where we average the error from the $f$ interfaces over the $n_i$ inner-code patches. See \aref{app:yoked_surface_codes_with_interfaces} for more details on the derivation of this formula and the fitting results for different yoked surface codes with interfaces.

Collectively, these QEC-level adaptations absorb the primary performance penalties of hardware modularity and eliminate the resulting complexity in architectural design. This enables a direct evaluation of the system-level spacetime overhead in utility-scale algorithms.

\begin{figure*}[htbp]
    \centering
    \includegraphics[width=0.9\linewidth]{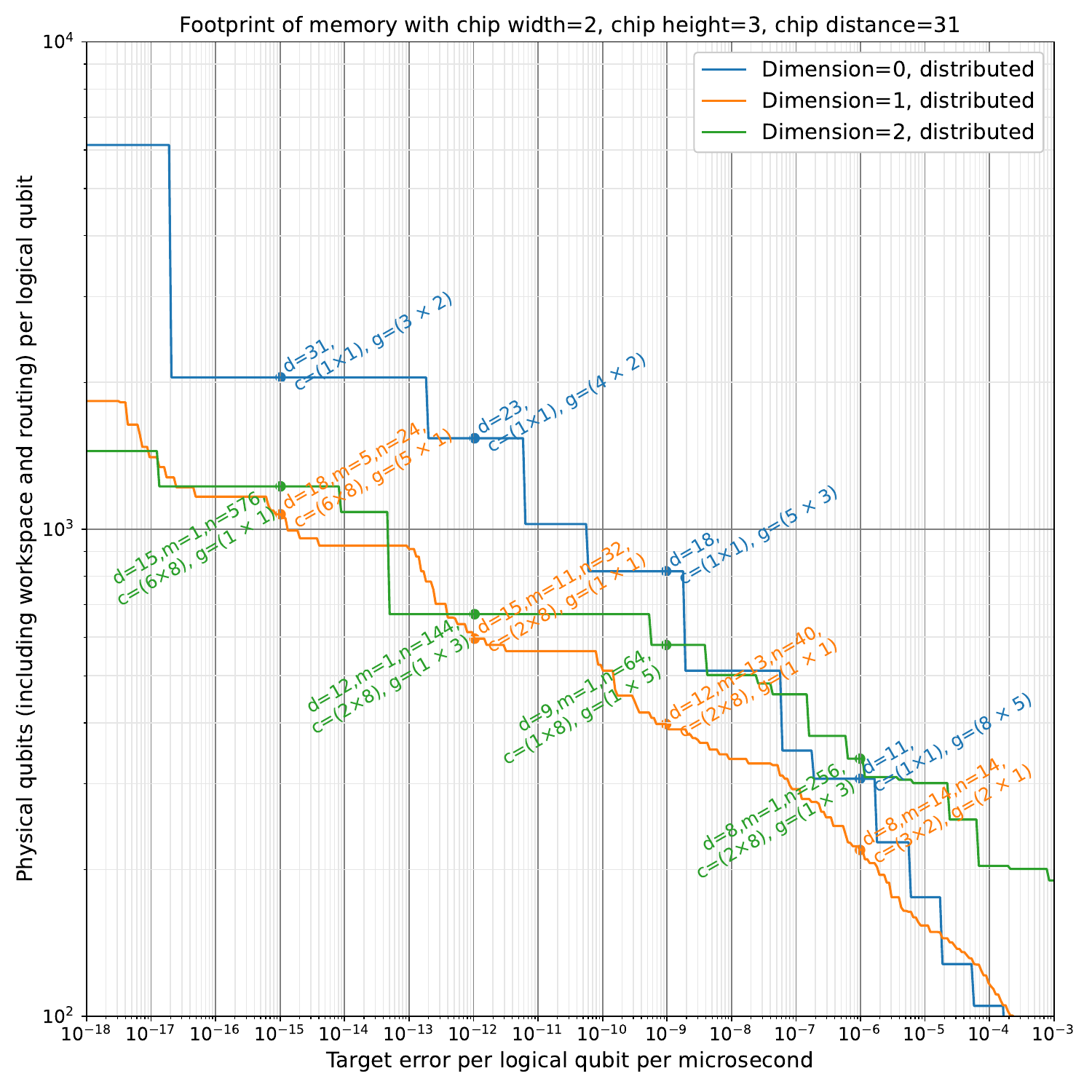}
    \caption{Inverse code rates of the standard surface code (Dimension=0) and yoked surface codes (Dimension=1, 2) for memory in the distributed architecture under varied target logical error rates. The chip size here is $h=3$, $w=2$, $d=31$, and latency-decoupled protocol is used. The labels $d$, $m$, $n$, $c$, and $g$ indicate the surface-code patch diameter ($d$), the number of \emph{CodeBlock}s per \emph{CodeGroup} ($m$), the number of used surface-code patches per \emph{CodeBlock} ($n$), the layout of chips for each \emph{CodeLayout} ($c$), and the layout of \emph{CodeGroup}s in each \emph{CodeLayout} ($g$).}\label{fig:footprint}
\end{figure*}

\section{Case study: RSA-2048 factorization}\label{sec:factorization_resource_estimation}

As a case study, we evaluate the resource overhead for factoring 2048-bit RSA integers utilizing the procedure detailed in \sref{sec:resource_estimation}.
We adopt the RSA-2048 implementation of Ref.~\cite{gidney_how_2025}, which is designed for high space efficiency by combining approximate residue arithmetic with a compact fault-tolerant layout.
At the algorithmic level, the dominant fault-tolerant workload reduces to additions, lookups, and phaseup operations.
Our architectural analysis uses these operation counts and the associated logical circuits as inputs, and can be applied to other realizations of modular exponentiation after substituting their corresponding counts and circuits.

We map the three functional regions of the monolithic architecture in Ref.~\cite{gidney_how_2025} onto the two-zone resource model introduced above.
The computation zone performs $1.1\times 10^7$ additions, $7.6\times 10^6$ lookups, and $1.6\times 10^6$ phaseup operations, and it stores the 131 active logical qubits in standard surface-code patches.
It consumes CCZ states produced by 6 factories, each performing magic-state cultivation~\cite{gidney_magic_2024} followed by 8T-to-CCZ distillation~\cite{gidney_efficient_2019}.
Together, the computation zone occupies $18\times 15$ logical patches.
The memory zone uses yoked surface codes to store the 1280 mostly idle input qubits at higher density. Thus, this instance is representative in both problem and architecture.

In the distributed architecture, the inter-chip error, extra latency, and finite chip capacity must be taken into account. The syndrome-extraction circuit in \fref{fig:interchip_circuit} isolates noisy, slow inter-chip operations from the intra-chip QEC cycle, so modularity changes the logical-error and timing parameters without requiring a different algorithmic schedule or substantially greater logical-level design complexity.

For the chip-level scan, each chip is parameterized by $(h,w,d)$, where $h$ and $w$ are the numbers of logical patches along the two chip directions and $d$ is the surface-code distance. For the latency-decoupled protocol, the corresponding chip capacity is
\begin{align}
    C_{\mathrm{chip}}&=2hw(d+1)^2+N_{\mathrm{int}}-1, \nonumber\\
    N_{\mathrm{int}}&=3(h+w)[2(d+1)-1],
\end{align}
where $N_{\mathrm{int}}$ counts the extra interface measurement qubits and buffer qubits. For simplicity, we use a uniform chip size across both zones, but the procedure can be extended to independent chip sizes for computation and memory.

For the \emph{prerequisite inputs}, we use the circuit-level model described in \sref{sec:circuit_and_error_modeling}. The physical error model sets the intra-chip bulk error rate to $p_b=0.1\%$ and inter-chip error rate $p_s=\gamma p_b=1\%$ with $\gamma=10$; the intra-chip two-qubit gate duration is assumed to be $40~\text{ns}$, and the inter-chip operation duration is $\eta \times 40$ ns. The QEC specification uses surface code syndrome extraction with a $1~\mu\text{s}$ cycle time, the latency-decoupled interface circuit of \fref{fig:interchip_circuit}, and the fitted logical-error ansatz of \eref{eq:ler_ansatz}. The control-system reaction time is taken to be $10~\mu\text{s}$, and the memory-zone estimates additionally use the yoked-surface-code formalism in \eref{eq:yoke_ler}. For the algorithm, the subroutines and the number of logical qubits remain the same as in the monolithic case, and we set the overall target algorithmic logical error rate to 10\%.

\begin{table}
    \centering
    \caption{Runtime estimates for the dominant RSA-2048 subroutines. Each row specifies the code distance, interface protocol, and inter-chip time factor used to estimate the time of one addition, lookup, and phaseup operation; the final column gives the one-shot runtime. Here, L-decoupled and L-dependent denote the latency-decoupled and latency-dependent interface protocols, respectively. The N/A row gives the monolithic comparison case. For a controlled comparison, we do not use rounded-up numbers as in~\cite{gidney_how_2025}, and as a result, the total runtime for the monolithic case is lower than the rounded-up numbers by about 25\%~\cite{gidney_answer_2025}. The distributed architecture rows take $h=w=3$ as an example.}
    \label{tab:rsa2048_subroutine_runtimes}
    \renewcommand{\arraystretch}{1.3}
    \begin{tabular*}{\linewidth}{@{\extracolsep{\fill}}|c|c|c|c|c|c|c|}
        \hline Protocol & $d$  & $\eta$ & Addition & Lookup & Phaseup & Total
        \\\hline  Monolithic & 27 & N/A & $1458\,\mu\mathrm{s}$ & $1953\,\mu\mathrm{s}$ & $635\,\mu\mathrm{s}$ & 8.7 h
         \\\hline {\bf L-decoupled}  & {\bf 31} & $\le 25$ &  $\boldsymbol{1875\,\mu\mathrm{s}}$ & $\boldsymbol{2495\,\mu\mathrm{s}}$ & $\boldsymbol{788\,\mu\mathrm{s}}$ & {\bf 11.2 h} 
        \\\hline  L-dependent & 31 & 5 & $2009\,\mu\mathrm{s}$ & $2671\,\mu\mathrm{s}$ & $840\,\mu\mathrm{s}$ & 12.0 h
        \\\hline  L-dependent & 31 & 10 & $2344\,\mu\mathrm{s}$ & $3111\,\mu\mathrm{s}$ & $971\,\mu\mathrm{s}$ & 13.9 h
        \\\hline  L-dependent & 31 & 15 & $2678\,\mu\mathrm{s}$ & $3551\,\mu\mathrm{s}$ & $1101\,\mu\mathrm{s}$ & 15.9 h
        \\\hline  L-dependent & 33 & 20 & $3208\,\mu\mathrm{s}$ & $4245\,\mu\mathrm{s}$ & $1303\,\mu\mathrm{s}$ & 19.1 h
        \\\hline   L-dependent & 33 & 25 & $3564\,\mu\mathrm{s}$ & $4713\,\mu\mathrm{s}$ & $1442\,\mu\mathrm{s}$ & 21.2 h
        \\\hline
    \end{tabular*}
\end{table}

\begin{table}
    \centering
    \caption{Logical error estimates for the dominant RSA-2048 subroutines. Each row uses the same architecture parameters as \tref{tab:rsa2048_subroutine_runtimes}. The Addition, Lookup, and Phaseup columns give the logical error contribution of one operation, and the final column gives the aggregate logical error contribution.}
    \label{tab:rsa2048_subroutine_errors}
    \renewcommand{\arraystretch}{1.3}
    \begin{tabular}{|c|c|c|c|c|c|c|}
        \hline   Protocol                   &$d$ &  $\eta$ & Addition & Lookup & Phaseup & Total
        \\\hline Monolithic                        & 27 &  N/A & $7.5\times 10^{-10}$ & $1.0\times 10^{-9}$ & $3.3\times 10^{-10}$ & 1.61\% 
        \\\hline {\bf L-decoupled}    & 31 &  $\le 25$ & $8.0\times 10^{-10}$ & $1.1\times 10^{-9}$ & $3.4\times 10^{-10}$ & 1.71\% 
        \\\hline L-dependent            & 31 &  5 & $1.1\times 10^{-9}$ & $1.5\times 10^{-9}$ & $4.6\times 10^{-10}$ & 2.33\% 
        \\\hline L-dependent            & 31 &  10 & $2.3\times 10^{-9}$ & $3.0\times 10^{-9}$ & $9.5\times 10^{-10}$ & 4.79\% 
        \\\hline L-dependent            & 31 &  15 & $4.8\times 10^{-9}$ & $6.4\times 10^{-9}$ & $2\times 10^{-9}$ & 9.78\% 
        \\\hline L-dependent            & 33 &  20 & $1.6\times 10^{-9}$ & $2.1\times 10^{-9}$ & $6.5\times 10^{-10}$ & 3.34\% 
        \\\hline L-dependent            & 33 &  25 & $3.1\times 10^{-9}$ & $4.1\times 10^{-9}$ & $1.3\times 10^{-9}$ & 6.40\% 
        \\\hline
    \end{tabular}
\end{table}

For the \emph{computation} zone, we calculate the smallest code distance that meets the target logical error rate in order to minimize chip capacity. The number of chips required is $\lceil 18 / h\rceil \cdot \lceil 15/w \rceil$. For the shot time and the logical error rate, we conservatively estimate each subroutine's contribution by treating all inter-chip interfaces as persistently active throughout the execution of that subroutine, and then aggregate the contributions of all subroutines. The interface contribution to the error of each subroutine is calculated as the product of the total number of inter-chip interfaces, the number of rounds, and the error per interface ($p_L^s$). The bulk contribution is computed as the product of the number of logical qubits, the number of rounds, and the logical error rate per qubit. As an example, the resulting subroutine runtimes and logical-error estimates for the $h=w=3$ case are shown in \tref{tab:rsa2048_subroutine_runtimes} and \tref{tab:rsa2048_subroutine_errors}, respectively. If the aggregate logical error rate exceeds the target, we incrementally increase the code distance and repeat the estimation until the requirement is met. Note that a more precise estimation requires analyzing the number of inter-chip lattice surgeries for each specific chip geometry, and we provide a detailed analysis in \aref{app:detailed_runtime_analysis}.

Then we optimize the parameters in the \emph{memory} zone. Given the logical error rate $\text{LER}_c$ and shot time obtained from the computation zone, total target logical error rate $\text{LER}^{\text{tar}}$, and the number of logical qubits to be stored, we can determine the target logical error rate per qubit per microsecond for the yoked surface code. For each target logical error rate, we optimize the code parameters and layouts by grid scanning to maximize the code rate, defined as the ratio of encoded logical qubits to physical qubits across all memory chips. An example for $h=3, w=2, d=31$ is shown in \fref{fig:footprint}. Details of the optimization of yoked surface codes are provided in \aref{app:yoked_surface_codes_with_interfaces}. The final code parameters and layout are selected by first maximizing the code rate and then minimizing the total chip count.

We aggregate resource requirements. Representative estimates for square chip geometries are shown in \fref{fig:spacetime_overhead_1e-3}, and more detailed data are summarized in \tref{tab:rsa2048_resource_estimates}.

\begin{table}[h]
\centering
\caption{Resource estimates for RSA-2048 factorization under different square chip geometries. The rows with $(h,w,d)$ list distributed architectures, where $h$ and $w$ are the numbers of logical qubits along the height and width of each chip, $d$ is the code distance of each patch, $C_{\mathrm{chip}}$ is the physical qubit count per chip, $N_{\mathrm{chip}}$ is the number of chips, $N_{\mathrm{total}}$ and $N_{\mathrm{used}}$ are the total and actively used physical qubit counts, and $T_{\mathrm{total}}$ is the expected total runtime. The N/A row gives the monolithic comparison case. Note that the distance for the monolithic case is 27 instead of 25 as reported in \cite{gidney_how_2025} because we use the SI1000 noise model for it.}
\label{tab:rsa2048_resource_estimates}
    \renewcommand{\arraystretch}{1.3}
    \begin{tabular}{|c|c|c|c|c|c|}
        \hline $(h,w,d)$ & $C_{\mathrm{chip}}$ & $N_{\mathrm{chip}}$ & $N_{\mathrm{total}}$ & $N_{\mathrm{used}}$ & $T_{\mathrm{total}}$ (days)
        \\\hline (1,1,31) & 2425 & 837 & 2.03e+06 & 1.50e+06 & 4.42 
        \\\hline (2,2,31) & 8947 & 252 & 2.25e+06 & 1.50e+06 & 4.39 
        \\\hline (3,3,31) & 19565 & 93 & 1.82e+06 & 1.50e+06 & 4.37 
        \\\hline (4,4,31) & 34279 & 65 & 2.23e+06 & 1.39e+06 & 4.63 
        \\\hline (5,5,31) & 53089 & 36 & 1.91e+06 & 1.68e+06 & 4.48 
        \\\hline (6,6,31) & 75995 & 23 & 1.75e+06 & 1.39e+06 & 4.50 
        \\\hline (7,7,31) & 102997 & 18 & 1.85e+06 & 1.42e+06 & 4.39
        \\\hline (8,8,31) & 134095 & 13 & 1.74e+06 & 1.38e+06 & 4.38 
        \\\hline (9,9,31) & 169289 & 11 & 1.86e+06 & 1.63e+06 & 4.36 
        \\\hline (N/A, N/A, 27) & $1.26\times 10^6$ & 1 & 1.26e+06 & 1.26e+06 & 3.40 
        \\\hline
    \end{tabular}
\end{table}

The space overheads corresponding to various chip sizes are shown in \tref{tab:rsa2048_resource_estimates} and \fref{fig:spacetime_overhead_1e-3}. In \fref{fig:spacetime_overhead_1e-3}, the left panel shows the required number of chips as a function of chip capacity (qubits per chip). The solid black line indicates the monolithic limit and effectively represents a constant total qubit count. Colored lines correspond to distributed architectures with different inter-chip operation time factors ($\eta$) and interface circuit protocols. The data points on each line represent nine square chip configurations with $1\le h=w\le 9$. The results show that the extra space overhead is both modest and nearly invariant.
Variations in chip size have little impact on the total qubit count in the distributed architecture. As shown in \tref{tab:rsa2048_resource_estimates}, the actually used physical qubit count is smaller than the total allocated count: across the square-chip configurations, only $62\%$--$88\%$ of the allocated physical qubits are used, leaving the rest qubits as layout slack. Therefore, the reported total qubit counts are conservative, and further optimization could improve qubit utilization. While the space overhead of the latency-dependent protocol increases with $\eta$, the architecture using the latency-decoupled protocol maintains a constant overhead, and becomes more space-efficient than the latency-dependent approach for large $\eta$.

The time overhead is shown in the right panel of \fref{fig:spacetime_overhead_1e-3}, where the total runtime of the entire algorithm is plotted as a function of $\eta$. We derive this runtime by multiplying the single-shot duration by the expected number of repetitions (9.2) and dividing by the single-shot fidelity. The black star at $\eta=0$ indicates the runtime of the monolithic architecture. The blue line depicts the average runtime for the latency-dependent protocol, which increases almost linearly with $\eta$. In contrast, the green line shows the latency-decoupled protocol, for which the runtime remains constant up to $\eta=25$. Shaded areas denote the upper and lower bounds across all searched configurations. Compared with the latency-dependent protocol, the latency-decoupled protocol substantially mitigates the temporal overhead induced by slower inter-chip operations and keeps the total runtime within a factor of 1.5 of the monolithic architecture.

The extra spacetime overhead is nearly scale-invariant even as the chip capacity is scaled across orders of a broad range, from few thousands to hundreds of thousands physical qubits per QPU. 
This capacity window is also well matched to industrial 6-inch and 12-inch wafer fabrication for superconducting qubits.
The weak dependence on chip capacity decouples individual chip size from overall system performance. This scale invariance turns chip capacity from a fine-tuned algorithmic parameter into a robust engineering degree of freedom. On the system level, the distributed architecture no longer requires a separate modularity-driven optimization strategy for each algorithm or chip size. On the physical level, chip-size choices can instead be guided more by fabrication yield, control and packaging complexity, and incremental system scaling than by a sharp penalty in end-to-end fault-tolerant cost.

\begin{figure}[t]
    \centering
    \includegraphics[width=\linewidth]{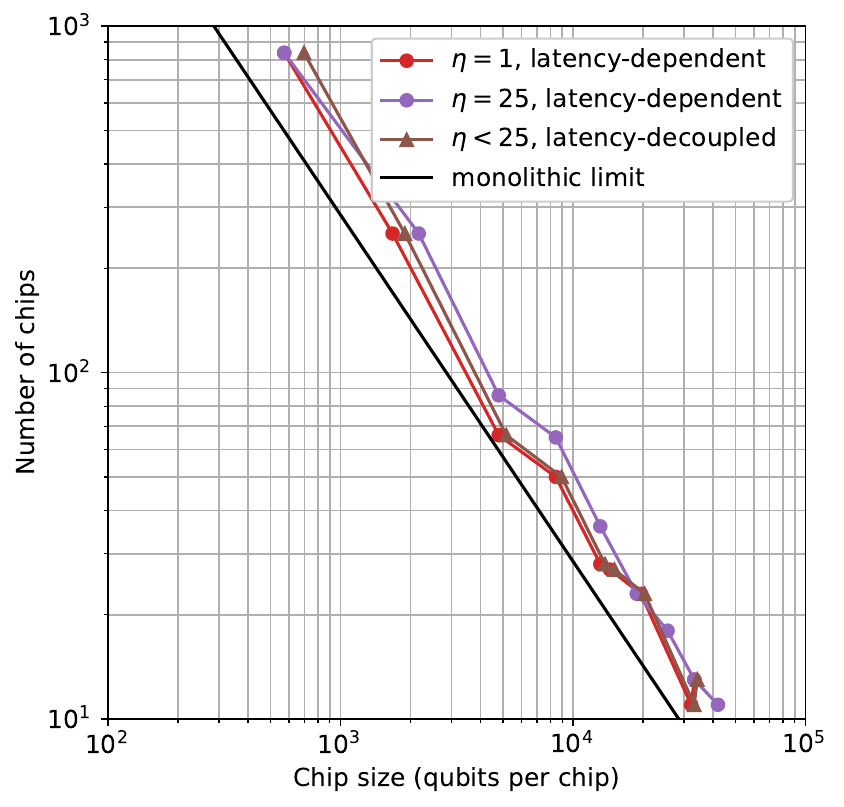}
    \caption{Space overhead of distributed architectures for factoring 2048-bit RSA integers, with physical error rate $p_b=10^{-4}$ and $p_s=10^{-3}$, using the same plotting conventions as \fref{fig:spacetime_overhead_1e-3}. The total distributed qubit cost remains nearly invariant across different chip sizes, decoupling chip size from global performance.}
    \label{fig:space_overhead_1e-4}
\end{figure}

\begin{figure}[t]
    \centering
    \includegraphics[width=\linewidth]{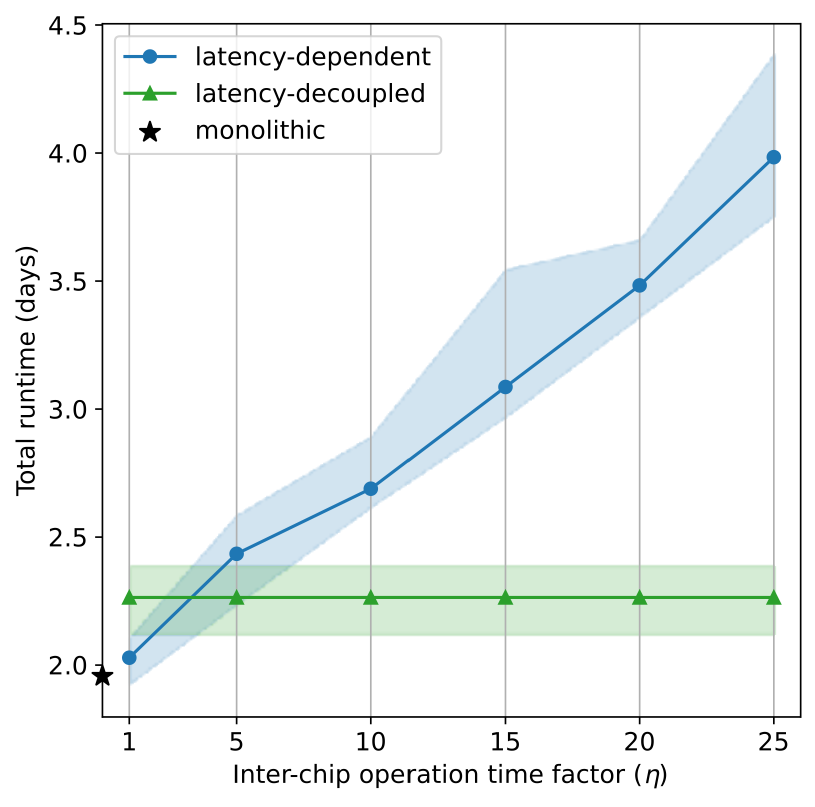}
    \caption{Time overhead of distributed architectures for factoring 2048-bit RSA integers, with physical error rate $p_b=10^{-4}$ and $p_s=10^{-3}$, using the same plotting conventions as \fref{fig:spacetime_overhead_1e-3}. The latency-decoupled protocol keeps the runtime independent of $\eta$, decoupling the runtime from the latency of the inter-chip operations.}\label{fig:time_overhead_1e-4}
\end{figure}

So far, we have focused primarily on a physical error rate of $p_b=0.1\%$ and $p_s=1\%$ because such an error rate is plausible to near-term superconducting hardware. As quantum hardware continues to improve, lower physical error rates may become achievable in the future. We also estimate the space-time overhead at $p_b=0.01\%$ and $p_s=0.1\%$. In this regime, the computation-zone distance is roughly halved, reducing the chip capacity by about a factor of four relative to the $p_b=0.1\%$ baseline. Magic state cultivation also becomes cheaper: using Fig.~29 of~\cite{gidney_magic_2024}, we reduce $d_1$ from 5 to 3 and use an expected cultivation volume below $5\cdot 10^3$ for a $|T\rangle$ state with error below $10^{-7}$. For memory, we estimate the lower logical error rate $p_{c,k}$ in \eref{eq:yoke_ler} from the fits at $p_b=0.1\%$, rather than from full low-error circuit simulations; details are given in \aref{app:yoked_surface_codes_with_interfaces}. The resulting space-time costs, shown in \fref{fig:space_overhead_1e-4} and \fref{fig:time_overhead_1e-4}, preserve the main conclusion: the extra space-time distributed overhead remains modest and nearly scale-invariant across chip sizes of a broad range.

\section{Conclusion and outlook}\label{sec:conclusion}

In this work, we show that distributed superconducting fault-tolerant quantum computation can approach monolithic performance under conservative, experimentally anchored hardware constraints, with overhead that depends only weakly on chip size. Modularity therefore need not impose prohibitive spacetime cost or demand heavy system-level orchestration. We substantiate this conclusion through a hardware-grounded architectural co-design and a resource-estimation protocol that integrates superconducting hardware constraints, circuit-level error-correction simulations, and the workload of a utility-scale algorithm.

The near scale-invariant behavior is enabled by absorbing modularity-induced performance penalties at module boundaries. Through QEC-level adaptations, the interface design shields bulk stabilizer measurements from slow and noisy inter-chip operations, so that inter-chip latency and noise do not propagate into a system-wide performance bottleneck. This effectively decouples system performance from module size and removes chip capacity as an additional architectural optimization burden. 
Therefore, rather than being a fine-tuned parameter, chip capacity becomes a flexible engineering degree of freedom and can be chosen according to realistic constraints from manufacturability, packaging, and control.

We quantify this behavior using RSA-2048 factorization as a stringent benchmark, which provides a controlled comparison between distributed and monolithic architectures through established resource estimates. 
Under the baseline error model, with $p_b = 10^{-3}$ and inter-chip operations that are 10 times noisier and 25 times slower than local gates, the distributed architecture requires about 2.0 million physical qubits and 4.4 days. This remains close to the corresponding monolithic baseline of 1.3 million physical qubits and 3.4 days, with only a modest increase in surface-code distance from $d = 27$ to $d = 31$. The same trend persists in the lower-error regime $p_b = 10^{-4}$, where the overhead remains modest and weakly dependent on module capacity. This result reflects an architecture-level feature of surface-code-based superconducting computation rather than a task-specific effect. In such architectures, cross-module logical operations are implemented through Pauli-measurement primitives whose underlying mechanism is algorithm-agnostic.

Our estimates focus on surface-code-based superconducting architectures with planar layouts, nearest-neighbor coupling, and experiment-motivated physical parameters. This constrained setting deliberately reduces open assumptions and ties the resource estimates to experimentally accessible building blocks. Direct qubit-count comparisons with qLDPC-based proposals should therefore be interpreted with care, since those approaches may offer substantially lower overhead in principle but often rely on less settled assumptions about noise models, connectivity, decoding, logical operations, and hardware integration. The present focus is therefore timely, and the near scale-invariant overhead found here provides performance assurance for this experimentally relevant regime.

These results provide a foundation for future logical-level compilation and physical-layer integration in DFTQC. As the modularity overhead is localized at chip boundaries, many logical-level optimization strategies developed for monolithic surface-code architectures should remain applicable, with limited modification, to distributed processors. This reduces the need to redesign lattice-surgery schedules and spacetime routing around modularity-induced bottlenecks. A complementary direction is to incorporate physical-integration constraints, including interconnect layout, control wiring, and real-time decoding. This would further advance the ultimate goal of full-stack co-design for practical distributed fault-tolerant superconducting quantum computation.

\begin{acknowledgments}
This work is supported by the Strategic Priority Research Program of Chinese Academy of Sciences (Grant No. XDB1680000) and the National Natural Science Foundation of China (Grants No.~12675032).
\end{acknowledgments}

\section*{Data Availability}

The code and data will be made publicly available.

\appendix

\section{Noise model}
\label{app:noise_model}

Simulations in the main text were done using the distributed noise model with time-correlated idle errors. This noise model is parameterized by three values:
\begin{enumerate}
    \item $p_b$ (bulk error rate): The characteristic error rate for local single-qubit gates, two-qubit gates, resets and measurements within a single chip.
    \item $p_s$ (interface error rate): The error rate associated with the inter-chip coupling.
    \item $\eta$ (inter-chip operation time factor): A dimensionless multiplier accounting for the extended duration of inter-chip operations relative to local gates.
\end{enumerate}

We define the Pauli noise channels $\mathcal{E}_P(\rho, p)$ ($P\in \{X, Y, Z\}$) and standard single-qubit and two-qubit depolarizing channels, $\mathcal{E}_1(\rho, p)$ and $\mathcal{E}_2(\rho, p)$, as follows:
\begin{equation}
    \mathcal{E}_P(\rho, p) = (1-p)\rho + pP\rho P, \quad P\in \{X, Y, Z\}
\end{equation}
\begin{equation}
    \mathcal{E}_1(\rho, p) = (1-p)\rho + \frac{p}{3}(X\rho X + Y\rho Y + Z\rho Z)
\end{equation}
\begin{equation}
    \mathcal{E}_2(\rho, p) = (1-p)\rho + \frac{p}{15}\sum_{P \in \{I,X,Y,Z\}^{\otimes 2} \setminus \{I\otimes I\}} P \rho P
\end{equation}
We also denote the flip of measurement results as a classical bit-flip $\mathcal{E}_M(p)$ with probability $p$.

The application of these channels to specific circuit operations is detailed in \tref{tab:noise_model}. To investigate the effects of elevated idle error, we base our noise model on the SI1000 noise model which is more realistic for the superconducting system, instead of commonly used uniform depolarizing noise model. Unlike the error models in the monolithic architecture where idle errors are uniform, our model dynamically scales the idle error on idle qubits depending on the operation occurring elsewhere. During intra-chip operations, idling qubits undergo depolarization with probability $p_b/10$. However, during a time step when inter-chip operations occur, the extended operation time exposes all non-participating bulk qubits to higher decoherence. This is modeled by applying a depolarizing channel with probability $\eta \cdot p_b/10$.

\begin{table*}[htbp]
    \centering
    \caption{Summary of the distributed noise model with time-correlated idle errors. The model distinguishes local bulk operations from inter-chip operations and scales idle errors by $\eta$ during the latter.}
    \label{tab:noise_model}
    \renewcommand{\arraystretch}{1.3}
    \begin{tabular}{l l l c}
        \hline\hline
        \textbf{Operation} & \textbf{Target Qubits} & \textbf{Noise Channel} & \textbf{Error Probability} \\
        \hline
        Reset ($R_X$) & Active Qubit & $\mathcal{E}_Z(\rho, p)$ & $p = 2p_b$ \\
        Reset ($R_Z$) & Active Qubit & $\mathcal{E}_X(\rho, p)$ & $p = 2p_b$ \\
        1-Qubit Gate  & Active Qubit & $\mathcal{E}_1(\rho, p)$ & $p = p_b/10$ \\
        2-Qubit Gate (bulk) & Active Qubits & $\mathcal{E}_2(\rho, p)$ & $p = p_b$ \\
        2-Qubit Gate (inter-chip) & Active Qubits & $\mathcal{E}_2(\rho, p)$ & $p = p_s$ \\
        Measurement & Active Qubit & $\mathcal{E}_M(p_m)\cdot\mathcal{E}_1(\rho, p)$& $p = p_b, p_m=5p_b$ \\
        \hline
        Idle (during bulk ops) & Idle Qubits & $\mathcal{E}_1(\rho, p)$ & $p = p_b/10$ \\
        Idle (during interface ops) & Idle Qubits & $\mathcal{E}_1(\rho, p)$ & $p = \eta \cdot p_b/10$ \\
        \hline\hline
    \end{tabular}
\end{table*}

\section{Fitting methodology}\label{app:fitting_details}

For the benchmark, we use Stim as the sampler~\cite{gidney_stim_2021}, correlated matching as the decoder~\cite{oscarhiggott_oscarhiggottpymatching_2025}, and use the least-squares method to fit the data~\cite{newville_lmfit_2025,jacinto_network_2025}. Here we provide more details about the fitting procedure and results.

\subsection{Fitting procedure}

The ansatz for the logical X error rate per round takes the following form:
\begin{align}
    f(d,p_b,p_s;\{\theta\}) &= \alpha_1\left(\frac{p_{s}}{p_{s}^*}\right)^{\frac{d+1}{2}} +\alpha_2\left(\frac{p_b}{p_b^*}\right)^{\frac{d+1}{2}} \nonumber\\
    &+\alpha_3 \sum_{1 \leq i \leq d}\left(\frac{p_{s}}{p_{s}^*}\left[1+\frac{\alpha_c}{1-\sqrt{p_b/ p_b^*}}\right]\right)^{\frac{i}{2}}\left(\frac{p_b}{p_b^*}\right)^{\frac{d+1-i}{2}}, \label{eq:app_ler_ansatz}
\end{align}
where $\{\theta\} = \{\alpha_1, \alpha_2, \alpha_3, \alpha_c, p_b^*, p_s^*\}$ are the fitting parameters.

We conduct Monte-Carlo sampling of $|0\rangle_L$ memory experiments for each parameter set $(d, p_b, p_s, \eta)$ over $k = 3d$ rounds. Sampling continues until reaching either $10^3$ logical errors or $10^8$ shots, whichever occurs first. From the resulting data, we calculate the logical error rate per shot as $p_{L,k}(d, p_b, p_s, \eta) = \frac{\text{\# logical errors}}{\text{\# shots}}$ and subsequently extract the per-round logical error rate $p_{L}(d, p_b, p_s, \eta) =  1-(1-p_{L,k})^{1/k}$. The uncertainty estimate is given by:
\begin{equation}
\sigma_{p_L}=\sqrt{\frac{p_{L, k}\left(1-p_{L, k}\right)}{n}} /\left(k\left(1-p_{L, k}\right)^{1-1 / k}\right),
\end{equation}
where n denotes the total number of shots. The fitting procedure minimizes the chi-squared quantity:
\begin{equation}
\chi^2(\{\theta\})=\sum_{\left(d, p_b, p_s, \eta\right)}\left(\frac{p_L-f\left(d,p_b, p_s ;\{\theta\}\right)}{\sigma_{p_L}}\right)^2
\end{equation}

We perform the fit using a logarithmically-spaced grid of $(p_b, p_s)$ pairs covering the parameter space $p_b\in[10^{-4}, 10^{-3}]$ and $p_s\in[p_b, \min(50p_b, 5\times 10^{-2})]$. At each grid point, we sample across code distances $d \in \{3, 5, 7, 9, 11, 13\}$ but restrict the fit to $d \in \{5, 7, 9, 11, 13\}$ to avoid inaccuracies arising from the path counting approximation on small lattices. We further ensure data quality by retaining only points satisfying $\sigma_{p_L} < \frac{p_L}{2}$.

The latency-dependent interface protocol from Refs.~\cite{shalby_optimized_2025, jacinto_network_2025} only adds $2d-1$ ancilla qubits at the interface and requires bulk qubits to idle while waiting for the slower inter-chip operation to complete. This leads to an increased QEC cycle time of $(1000 + \eta \times 40)$ ns and accumulated idle errors. For this protocol, we set $\eta\in[1, 5, 10, 15, 20, 25]$.

For the latency-decoupled protocol, we use the same interface circuit as the latency-dependent protocol in simulation, because the errors of SWAP gates only slightly degrade the fidelity of inter-chip Bell pairs, which is negligible compared to the effects of $p_s$. Meanwhile, we account for the latency of the extra SWAP layer by setting $\eta=3$ effectively.

\subsection{Fitting results}

\begin{figure}[htbp]
    \centering
    \includegraphics[width=\linewidth]{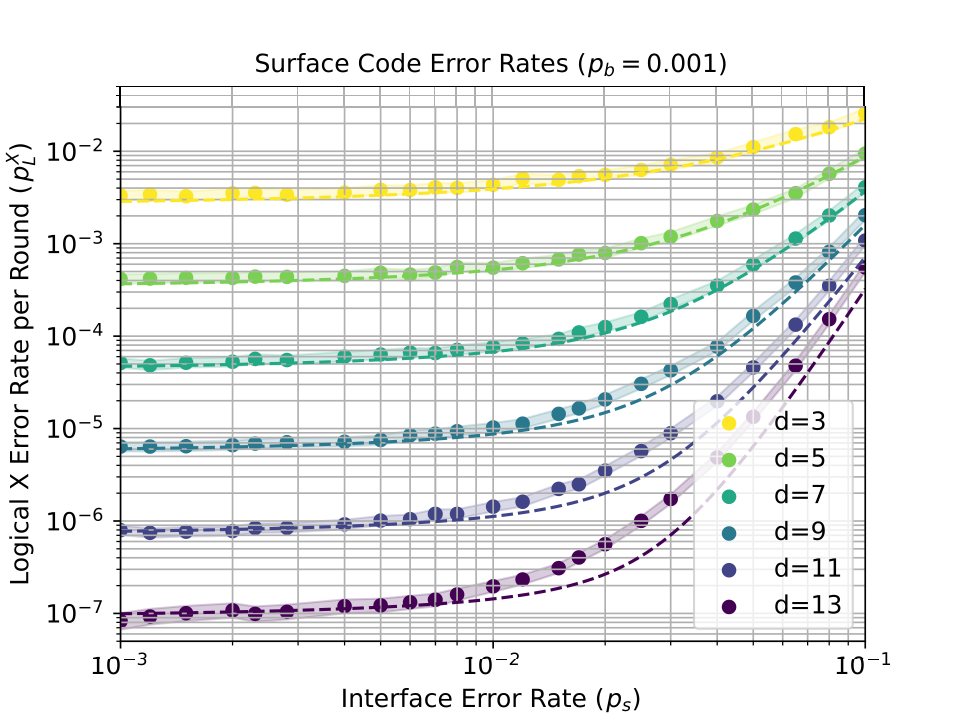}
    \includegraphics[width=\linewidth]{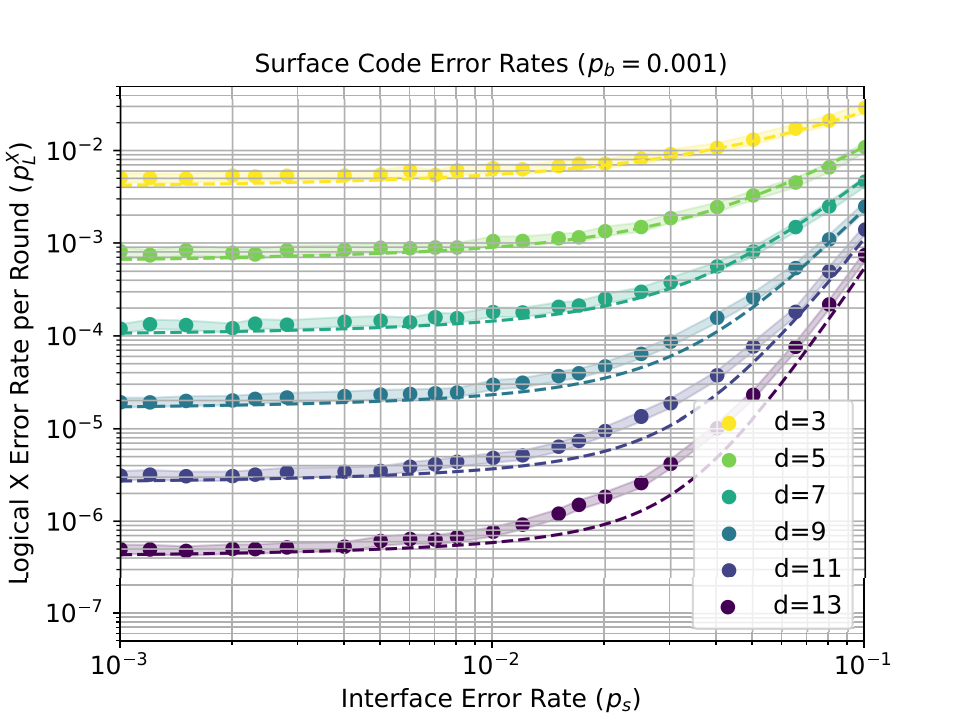}
    \caption{Logical X error rate per round $p_L^X$ as a function of the inter-chip physical error rate for various code distances under $\eta=1$ (top) and $\eta=25$ (bottom) for the latency-dependent protocol. The bulk physical error rate is fixed at $p_b = 10^{-3}$. Dashed curves show fits to the ansatz of \eref{eq:ler_ansatz}. Shading represents hypothetical logical error rates with a Bayes factor of at most 1000 relative to the maximum-likelihood hypothesis probability, assuming a binomial distribution. For $p_s>10p_b=10^{-2}$, the logical error rates deviate slightly from the fitting curves, but the main resource estimation results are unaffected because we focus on the case of $p_s = 10p_b$.}
    \label{fig:fitting_results}
\end{figure}

The fitting results for the latency-dependent protocol under $p_b= 10^{-3}$, $\eta=1$ and $\eta=25$ are shown in \fref{fig:fitting_results}. The fitting results for other $\eta$ values are similar and thus omitted here. Note that for $p_s>10p_b=10^{-2}$, the logical error rates deviate from the fitting curves slightly, but our main results of resource estimation are not affected since we focus on the case of $p_s = 10p_b$. The extracted fitting parameters are summarized in \tref{tab:fitting_results}.

\begin{table}[h]
\centering
\caption{Parameters of the ansatz in \eref{eq:app_ler_ansatz} for various values of $\eta$. Uncertainties are calculated as the square root of the diagonal elements of the covariance matrix via \texttt{scipy.optimize.leastsq}.}
\label{tab:fitting_results}
\begin{tabular}{|c|c|c|c|c|}
\hline
$\eta$ & $\alpha_1$        & $\alpha_2$        & $\alpha_3$         & $\alpha_c$        \\ \hline
1      & $0.100 \pm 0.004$ & $0.161 \pm 0.006$ & $0.0617 \pm 0.0024$ & $0.203 \pm 0.005$ \\ \hline
3      & $0.101 \pm 0.004$ & $0.162 \pm 0.006$ & $0.0597 \pm 0.0023$ & $0.212 \pm 0.005$ \\ \hline
5      & $0.103 \pm 0.004$ & $0.161 \pm 0.005$ & $0.0602 \pm 0.0023$ & $0.212 \pm 0.005$ \\ \hline
10     & $0.111 \pm 0.004$ & $0.160 \pm 0.005$ & $0.0599 \pm 0.0023$ & $0.220 \pm 0.005$ \\ \hline
15     & $0.100 \pm 0.004$ & $0.157 \pm 0.005$ & $0.0580 \pm 0.0023$ & $0.213 \pm 0.006$ \\ \hline
20     & $0.104 \pm 0.004$ & $0.155 \pm 0.005$ & $0.0560 \pm 0.0022$ & $0.220 \pm 0.006$ \\ \hline
25     & $0.112 \pm 0.004$ & $0.154 \pm 0.005$ & $0.0563 \pm 0.0023$ & $0.226 \pm 0.006$ \\ \hline
\end{tabular}

\vspace{1em}

\begin{tabular}{|c|c|c|}
\hline
$\eta$ & $p_s^*$           & $p_b^*$            \\ \hline
1      & $0.374 \pm 0.016$ & $0.00781 \pm 0.00005$ \\ \hline
3      & $0.375 \pm 0.017$ & $0.00767 \pm 0.00005$ \\ \hline
5      & $0.377 \pm 0.017$ & $0.00751 \pm 0.00005$ \\ \hline
10     & $0.383 \pm 0.017$ & $0.00717 \pm 0.00005$ \\ \hline
15     & $0.374 \pm 0.017$ & $0.00683 \pm 0.00004$ \\ \hline
20     & $0.377 \pm 0.017$ & $0.00653 \pm 0.00004$ \\ \hline
25     & $0.381 \pm 0.016$ & $0.00627 \pm 0.00004$ \\ \hline
\end{tabular}
\end{table}

\begin{figure}[!t]
    \centering
    \includegraphics[width=\linewidth]{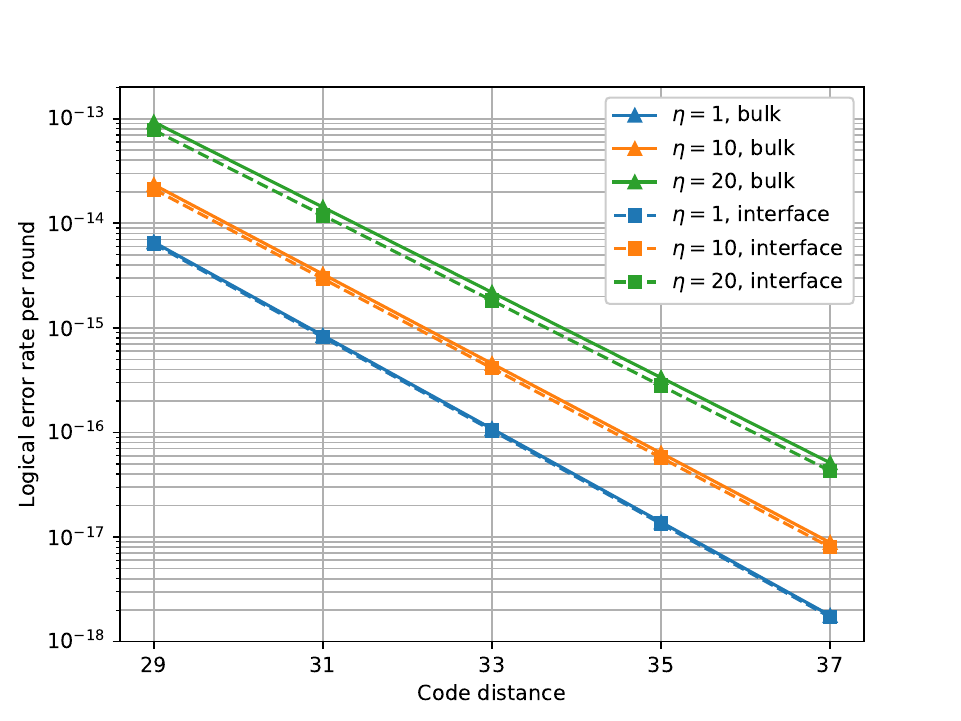}
    \caption{Logical error rates per round as functions of code distance under various inter-chip operation time factors ($\eta$) for the latency-dependent protocol. The bulk physical error rate is fixed at $p_b = 10^{-3}$ and the interface physical error rate is $p_s = 10 p_b$. Solid lines represent the bulk contribution ($p_L^b$), and dashed lines represent the interface contribution ($p_L^s$).}
    \label{fig:bulk_ler_vs_distance}
\end{figure}

The extrapolated logical error rates for the latency-dependent protocol are shown in \fref{fig:bulk_ler_vs_distance}. We display separately the bulk term $p_L^b=\alpha_2\left(\frac{p_b}{p_b^*}\right)^{\frac{d+1}{2}}$ and the interface-related terms $p_L^s=\alpha_1\left(\frac{p_{s}}{p_{s}^*}\right)^{\frac{d+1}{2}} +\alpha_3 \sum_{1 \leq i \leq d}\left(\frac{p_{s}}{p_{s}^{**}}\right)^{\frac{i}{2}}\left(\frac{p_b}{p_b^*}\right)^{\frac{d+1-i}{2}}$. We assume that the logical X and logical Z error rates are equal for a bulk patch, and because the simulation includes two patches, $p_L^b$ represents the logical error rate of an entire bulk patch. We see that achieving the same target logical error rate, such as $10^{-15}$ for Shor's algorithm, requires larger code distances at higher $\eta$ because of the increased idle errors. However, for the latency-decoupled protocol, the logical error rate is independent of $\eta$ in the regime we consider, so the required code distance remains constant. The advantage of the latency-decoupled protocol is therefore system-level: it moves the architectural penalty away from an $\eta$-dependent spacetime increase and reactive synchronization burden toward a modest interface-ancilla overhead.

\section{Subroutines of RSA integer factorization}\label{app:subroutines_of_rsa_factorization}

The factorization algorithm decomposes into three primary logical subroutines which dominate the space-time overhead: addition, lookup, and phaseup operations~\cite{gidney_how_2025}.

\textbf{Addition.} The addition operation performs the logical mapping $(a,b) \rightarrow (a, (b+a) \bmod 2^n)$ on an $n$-qubit offset register $a$ and an $n$-qubit target register $b$. It has a logical cost of $n-1$ Toffoli gates and requires $n-1$ ancilla qubits. The operation consists of two phases: computation, which performs an out-of-place addition producing the sum, and uncomputation, which removes one of the input registers (typically the previous accumulator value) using measurement-based uncomputation. The uncomputation creates phase corrections that are deferred and merged into later phaseup operations. In surface code implementations via lattice surgery, the operation exploits the structure of $Z$ parity products which are relatively inexpensive to access. Subtraction and comparison operations have nearly identical circuits and are counted as additions in cost estimation.

\textbf{Lookup.} The lookup operation, parameterized by a classical table $T$ of $2^n$ values, performs the mapping $(a, 0) \rightarrow (a, T_a)$ where $a$ is an $n$-qubit address register. It has a logical cost of $2^n - n - 1$ Toffoli gates and $n-1$ ancilla qubits. The implementation splits the address register into low and high halves, computes the power product of each half (a register storing all possible products of the original qubits), then performs multi-target Toffoli gates to initialize the output qubits. The power product computation uses "wandering" AND gates implemented via gate teleportation without immediate correction, with corrections handled by the classical control system. After the Toffolis complete, all ancillary qubits are measured in the $X$ basis, producing extensive phase corrections that are merged into a subsequent phaseup operation for efficient deferred handling.

\textbf{Phaseup.} The phaseup operation, parameterized by a classical bit table $T$ of size $2^n$, performs the mapping $\sum_k \alpha_k |k\rangle \rightarrow \sum_k (-1)^{T_k} \alpha_k |k\rangle$ on an $n$-qubit target register, negating amplitudes of computational basis states with non-zero table entries. It has a logical cost of approximately $2\sqrt{2^n}$ AND gates, $2\sqrt{2^n}$ workspace qubits, and $\sqrt{2^n}$ multi-target $CZ$ gates. This primitive is primarily required to handle phase corrections produced during measurement-based uncomputation of table lookups. The implementation follows a similar split-register strategy as lookup: the target register is divided into halves, each expanded into its power product, then masked phase flips are performed using multi-controlled $Z$ gates. The use of wandering AND gates reduces the reaction depth to $n/2 \pm O(1)$ instead of $n \pm O(1)$, as corrections only need to be applied layer-by-layer during the uncomputation phase.

\section{Detailed runtime analysis}
\label{app:detailed_runtime_analysis}
In this section, we present a detailed runtime analysis for a representative chip configuration with dimensions $h=3$ and $w=2$. We analyze the execution time of each subroutine, incorporating inter-chip operation overheads only where necessary. To isolate the impact of inter-chip operations, we adopt the latency-dependent protocol with $\eta=25$, resulting in a cycle time of $2\mu\text{s}$ for operations involving inter-chip communication. We assume physical error rates of $p_b=10^{-3}$ and $p_s=10p_b$, which necessitates a code distance of $d=33$. 

First, we analyze the runtime bottleneck. A lattice surgery layer without inter-chip operations takes $33\mu\text{s}$, whereas one involving inter-chip operations requires $2\cdot 33=66\mu\text{s}$. The larger code distance required by the distributed architecture provides additional space for magic state cultivation, thereby reducing its time cost for a fixed space-time volume. An 8T-to-CCZ factory spans $3\cdot 4\cdot 34^{2}\cdot 2$ qubits. The average time to cultivate a T state with a logical error rate of $10^{-7}$ is $\frac{8\times (3\times 10^4)}{3\cdot 4 \cdot 34^{2}\cdot 2}\approx 9$ rounds. In contrast, distillation time increases significantly due to the five layers of inter-chip lattice surgeries, requiring $2/3\cdot (33 + 5 \cdot 2\cdot 33)=242$ rounds when using temporally encoded lattice surgery~\cite{chamberland_universal_2022}. This totals $242+9\approx 251$ rounds per CCZ state. With 6 factories, the average CCZ production time is $42\mu\text{s}$. Consequently, the runtime is primarily limited by inter-chip lattice surgeries, followed by magic state production. Note that the 30,000 physical qubit $\cdot$ rounds space-time volume for cultivation is estimated under a uniform depolarizing noise model, but shifting to the SI1000 noise model hardly impacts the temporal overhead.

The most resource-intensive adders in the algorithm operate on two $n=33$ digit registers and proceed in two steps: compute and uncompute. In the compute step, 32 CCZ gates are executed. These are processed in parallel batches of 6, with each gate requiring approximately 6 layers of inter-chip lattice surgery. Therefore, the compute step takes $\lceil 32 /6\rceil \cdot 6\cdot 33\cdot 2\approx 2376\mu\text{s}$. In the uncompute step, processing occurs in batches of 6 digits, each completed in 3 lattice surgery layers, 2 of which cross chip interfaces. This costs $\lceil 32 /6\rceil \cdot (33+2\cdot 2\cdot 33)\approx 990\mu\text{s}$. In summary, the runtime of a 33-digit adder is approximately $3370 \mu\text{s}$.

The largest phaseup operations involve $n=6$ qubit addresses. We restrict these operations to using 3 factories to minimize inter-chip operations. Computing the "power product" of the half-registers uses 8 CCZ states, costing $\lceil 8/3\rceil \cdot 251=753\mu\text{s}$. Subsequently, 7 layers of inter-chip lattice surgeries are performed for the multi-target CZ gates, and 2 layers for uncomputation, costing $(2+7)\cdot 33\cdot 2= 594\mu\text{s}$. Accounting for one reaction layer after the AND gates and one layer before each uncomputation step, the total time is approximately $1380\mu\text{s}$.

The most complex lookups also use $n=6$ address qubits. A lookup is similarly based on the power product of the two half-registers but utilizes all 6 factories, meaning the runtime is dominated by inter-chip operations, taking $8\cdot 33\cdot 2=528\mu\text{s}$. Following this, 14 multi-target CX gates and 49 multi-target CCX gates are performed, both requiring inter-chip two-qubit gates, which costs $(14+49)\cdot 33\cdot 2 = 4158\mu\text{s}$. Including a final reaction delay layer, a lookup takes approximately $4700\mu\text{s}$ in total.

Based on these primitive runtimes, the one-shot runtime of the entire algorithm is approximately $1.1\cdot 10^7 \cdot 3370 \mu\text{s}+ 7.7 \cdot 10^6 \cdot 4700 \mu\text{s}+ 1.6\cdot 10^6 \cdot 1380 \mu\text{s}\approx 20.5$ hours.

If we overestimate the use of inter-chip operations as we did in \sref{sec:resource_estimation}, then the addition, phaseup, and lookup runtimes become $3570\mu\text{s}$, $1450\mu\text{s}$, and $4720\mu\text{s}$ respectively, leading to a total one-shot runtime of approximately $21.2$ hours. This indicates that our approximation introduces only a minor overestimation in the total runtime.

\section{Yoked surface codes with inter-chip operations}\label{app:yoked_surface_codes_with_interfaces}

\subsection{Review of yoked surface codes}

Yoked surface codes~\cite{gidney_yoked_2025} offer a method to significantly increase the logical qubit density of fault-tolerant memory without requiring non-planar connectivity. The core idea is to concatenate standard surface codes, which serve as the inner code providing high-fidelity physical qubits and lattice surgery capabilities, into a high-density parity check outer code. 

To measure the checks of the overlying code, extra workspace is required. In 1D yoked surface codes, we need a single workspace row, while in 2D case, there are a workspace row and a workspace column for each block. For clarification, we refer to an outer code with several inner code patches as a \emph{CodeBlock}, and a group of \emph{CodeBlock}s that shares the same workspace as a \emph{CodeGroup}. Thus a \emph{CodeGroup} of 1D yoked surface codes contains several \emph{CodeBlock}s arranged in a column, while a \emph{CodeGroup} of 2D yoked surface codes contains a single \emph{CodeBlock}, as shown in \fref{fig:CodeLayout}.

For a \emph{CodeGroup} of 1D yoked surface codes, the total length of a syndrome cycle scales as $r_i=8d_i\cdot (\text{\# of \emph{CodeBlock}s})+2d_i$, where $d_i$ is the code distance of the constituent inner surface code patches (also denoted as $d_m$ in \fref{fig:procedure}). For a \emph{CodeGroup} of $[[n^2, n^2-4n+2,4]]$ 2D yoked surface codes, the total length of a syndrome cycle scales as $r_i=25d_i\cdot\sqrt{n_i}+4d_i$.

The cumulative logical error rate of a \emph{CodeBlock} of $D=0,1,2$ yoked surface codes $p_{L,D}$ takes the following approximate form:
\begin{align}
    p_{L,0}&\approx n_i \cdot p_{c,0}, \nonumber\\
    p_{L,1}&\approx n_i^2 \cdot r_o \cdot p_{c,1}^2, \nonumber\\
    p_{L,2}&\approx n_i^2 \cdot r_o \cdot p_{c,2}^4,  \label{eq:yoke_ler_app}
\end{align}
where $r_o$ is the number of outer code rounds, $n_i$ is the number of inner code patches, and $d$ is the inner code distance.
$ p_{c,k} = r_i\cdot \frac{A_{k}^{-d}}{B_{k}}$ represents the probability that a length-$d$ error chain occurs in a single inner code patch during one round of the outer code. $A_{k}$ and $B_{k}$ are coefficients fitted from simulations under the SI1000 noise model.

\subsection{Yoked surface codes crossing one interface}\label{sec:yoked_surface_code_crossing_interface}

Next we analyze the effects of inter-chip operations on logical error rate of yoked surface codes. Our goal is to find out expressions similar to \eref{eq:yoke_ler_app}. In the first step, we consider the case with $\eta > 1$ and extra idle error is inserted. Then we add the extra logical error rate caused by the interface. Combining the two extra contributions, we obtain the approximate expressions to the cases where a \emph{CodeGroup} crosses one interface.

\begin{figure*}[htbp]
    \centering
    \resizebox{\linewidth}{!}{
        \includegraphics{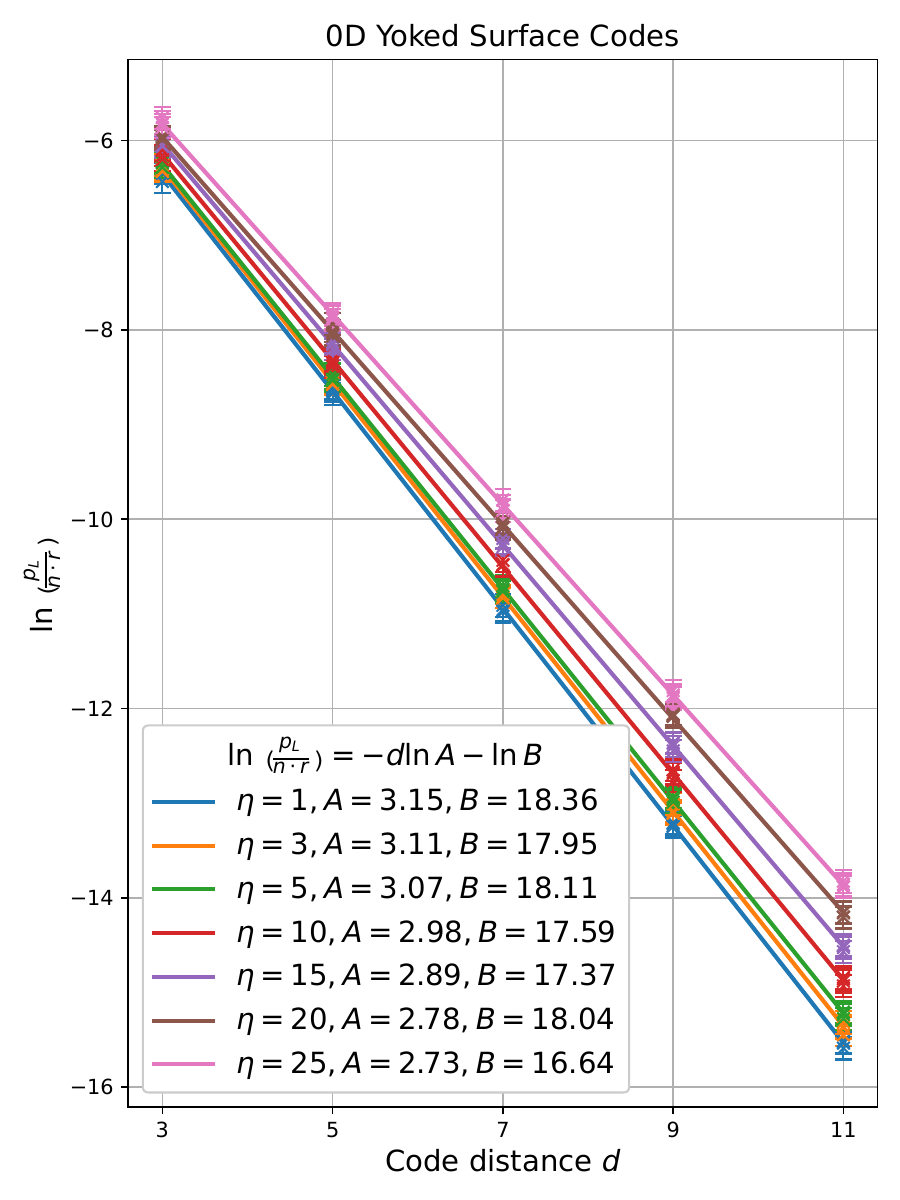}
        \hfill
        \includegraphics{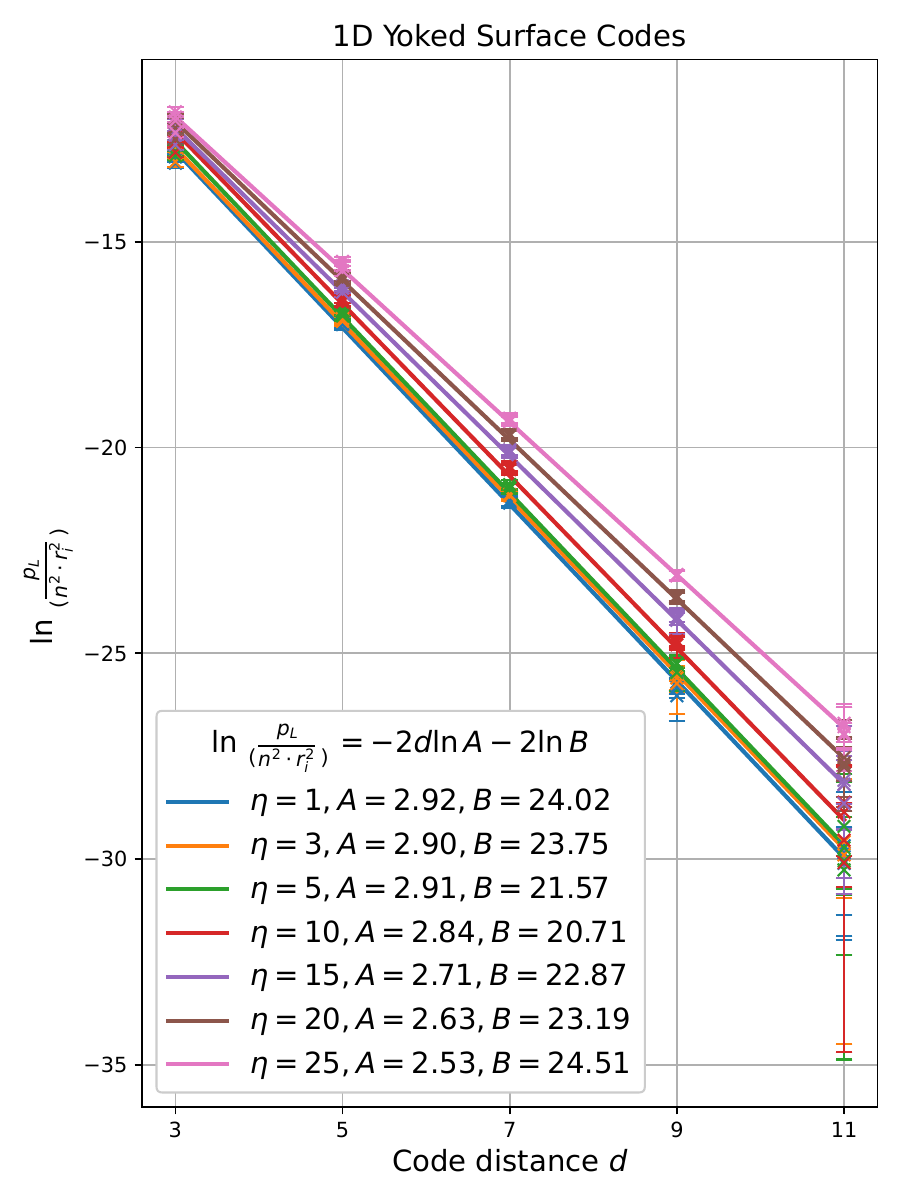}
        \hfill
        \includegraphics{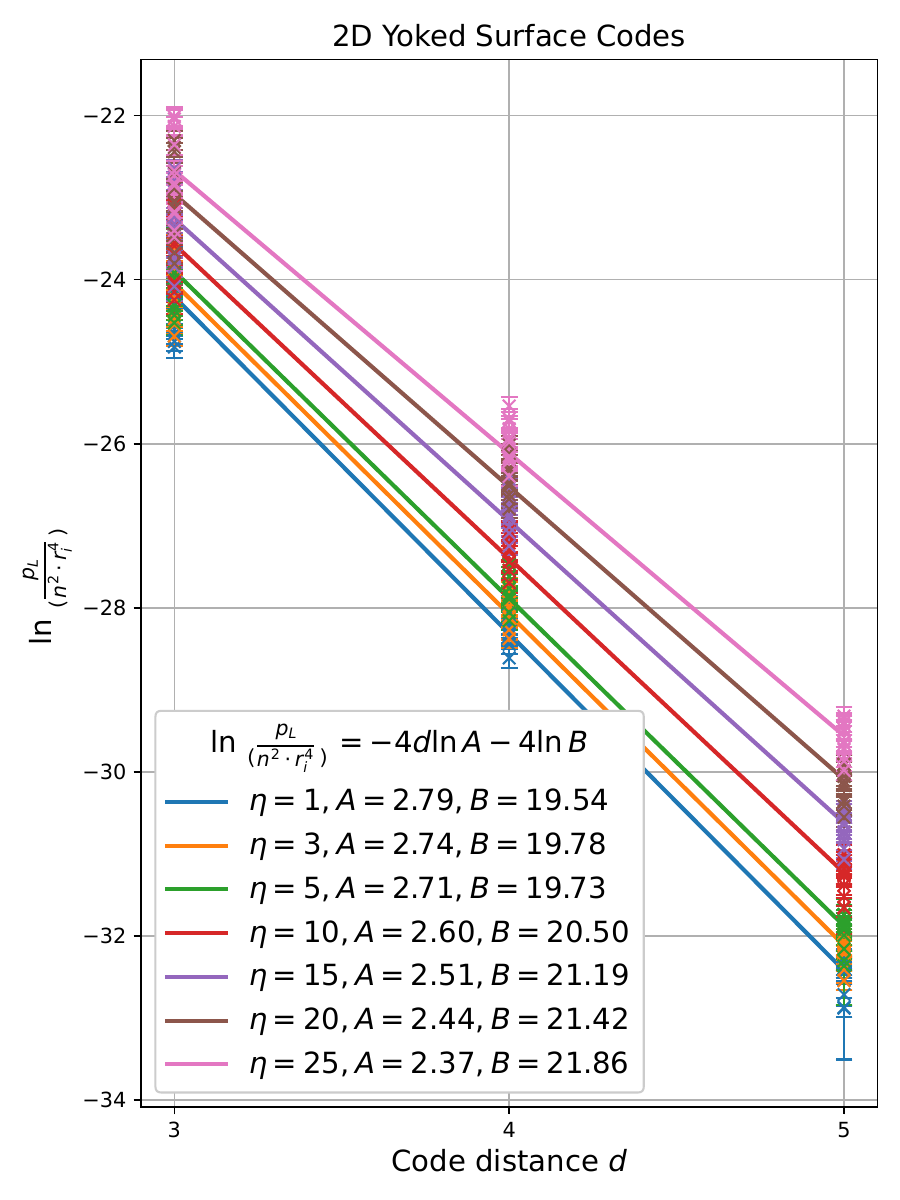}
    }
    \caption{Linear fits of logical error rates for yoked surface codes under different inter-chip operation time factors. Panels from left to right correspond to the 0D, 1D, and 2D cases, respectively. For the 0D and 1D cases, data with $r / d = 4, 8$ are used while encoding 4 or 8 logical qubits. For the 2D case, data with $r / d = 4, 8, 12$ and side lengths $n=4, 8, 12, 16$ are included.}\label{fig:fit_coef}
\end{figure*}

If extra idle error is inserted as $\eta>1$, \eref{eq:yoke_ler_app} still holds, but the fitting coefficients $A_{k}$ and $B_{k}$ should adjust correspondingly. In \fref{fig:fit_coef}, we show all fitting coefficients used in this work. If extra idle error is involved in $r_l$ rounds out of $r_i$ rounds, we can generalize the expression of $p_{c,k}$ as a linear combination of the $\eta=1$ and $\eta>1$ cases, i.e.
\begin{equation}
p_{c,k} = r_n\cdot \frac{A_{n,k}^{-d}}{B_{n,k}}+ r_l\cdot \frac{A_{l,k}^{-d}}{B_{l,k}},\label{eq:ler_for_mixed_noise_model}
\end{equation}
where the subscript $n$ and $l$ represent the coefficients fitted under $\eta=1$ and $\eta>1$ respectively, and $r_n=r_i - r_l$. Meanwhile, the total syndrome cycle time is $[(1+0.04\eta)\cdot r_l+r_n] \mu s$.

Next we add in the inter-chip operation errors at the interface. Generally the horizontal and vertical interfaces have different effects on the yoked surface codes. Suppose there are $r_{l,h}$ and $r_{l,v}$ inner code cycles with inter-chip two-qubit gates crossing horizontal and vertical interfaces respectively, and there are $f_h$ and $f_v$ horizontal and vertical length-d error chains at the interface respectively. Then the form of $p_{c,k}$ is modified to:
\begin{equation}
p_{c,k} = r_n\cdot \frac{A_{n,k}^{-d}}{B_{n,k}}+ r_l\cdot\frac{A_{l,k}^{-d}}{B_{l,k}} + \frac{f_h\cdot r_{l,h} + f_v\cdot r_{l,v}}{n_i}\cdot p_L^s,\label{eq:yoked_ler_with_interfaces}
\end{equation}
where we average the error of interfaces over $n_i$ patches. Here $p_L^s$ is the logical error rate per round of the interface as described in \sref{sec:circuit_and_error_modeling}.

With the expression of \eref{eq:yoked_ler_with_interfaces}, we consider the simple case where a \emph{CodeGroup} crosses only one interface, either horizontal or vertical.

For 1D yoked surface codes, if a \emph{CodeGroup} consisting of $[[n,n-2,2]]$ \emph{CodeBlock}s crosses a horizontal interface, the parity check of one \emph{CodeBlock} next to the interface will involve inter-chip coupling between that \emph{CodeBlock} and the workspace row. This affects $4d_i$ rounds of original stabilizer checks. Meanwhile, the $2d_i$ rounds of walking surface code~\cite{mcewen_relaxing_2023} are also affected. Therefore, $r_{l,h}=4d_i+2d_i=6d_i$ and $r_n=r_i-r_{l,h}=8d_i\cdot (\text{\# of \emph{CodeBlock}s}) -4d_i$. For one \emph{CodeBlock} next to the interface, there are $f_h=n$ error paths of length $d_i$, while for the others, $f_h=0$.

If a \emph{CodeGroup} of 1D crosses a vertical interface, then $f_v=1$. Each \emph{CodeBlock} will cross the interface once during its parity check with $4d_i$ rounds, so $r_{l,v}=4d_i\cdot (\text{\# of \emph{CodeBlock}s})$, $r_n=r_i-r_{l,v}=4d_i\cdot (\text{\# of \emph{CodeBlock}s}) + 2d_i$. 

For a \emph{CodeGroup} of $[[n^2, n^2-4n+2,4]]$ 2D yoked surface codes, the effects of crossing a horizontal interface are the same as crossing a vertical interface due to its symmetry in space. For example, crossing a horizontal interface will lead to $r_{l,h}=8d_i\cdot\sqrt{n_i}+6d_i$ rounds with inter-chip couplings, where $2d_i$ rounds are from walking surface codes, $4d_i$ rounds are from row and birow checks, and the rest $8d_i \cdot \sqrt{n_i}$ rounds are from column and bicolumn checks. Meanwhile, $f_h=\sqrt{n_i}$ and $r_n = r_i-r_{l,h} = 17 d_i \cdot \sqrt{n_i} -2 d_i$. Thus the expression of $p_{c,2}$ is:
\begin{equation*}
p_{c,2} = r_n\cdot \frac{A_{n,2}^{-d}}{B_{n,2}}+ r_{l,h}\cdot\frac{A_{n,2}^{-d}}{B_{n,2}} + r_{l,h}\cdot \sqrt{n_i}\cdot\frac{A_{l,2}^{-d}}{B_{l,2}}.
\end{equation*}

\subsection{Realistic layout with multiple chips}

\begin{figure*}[htbp]
    \centering
    \includegraphics[width=\linewidth]{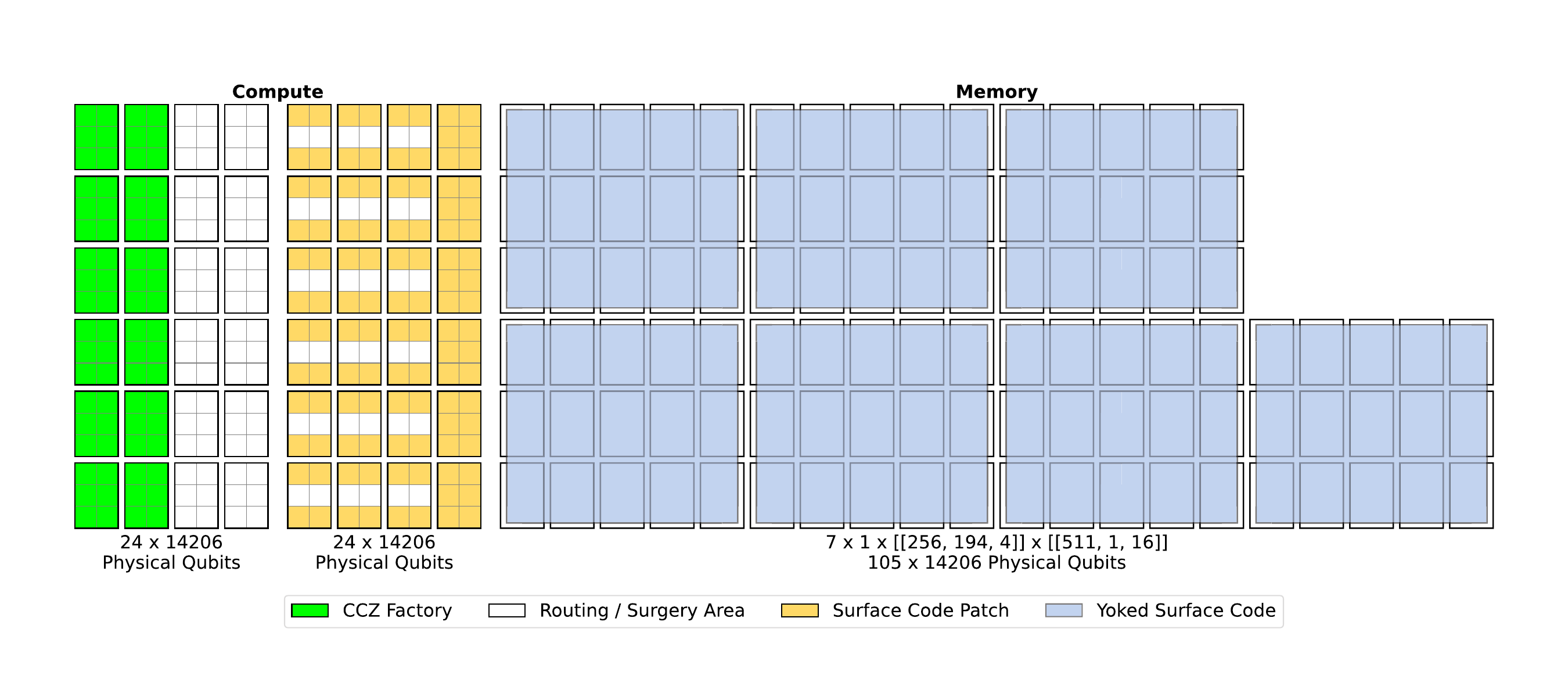}
    \caption{Optimized layout in the DFTQC architecture for chip size $h=3$, $w=2$, $d=33$ under the latency-dependent protocol with $\eta=25$. The compute zone uses 24 chips each for active compute and hot storage, while the memory zone uses 105 chips, for a total of 154 chips. Each chip has $1.42\times 10^4$ qubits, summing to $2.2$ million physical qubits. The memory uses 7 \emph{CodeLayout}s, each with $1\times [[256, 194, 4]]$ blocks and inner code distance 16.}\label{fig:space_layout_demonstration}
\end{figure*}

\begin{figure*}[htbp]
    \centering
    \resizebox{\linewidth}{!}{
        \includegraphics{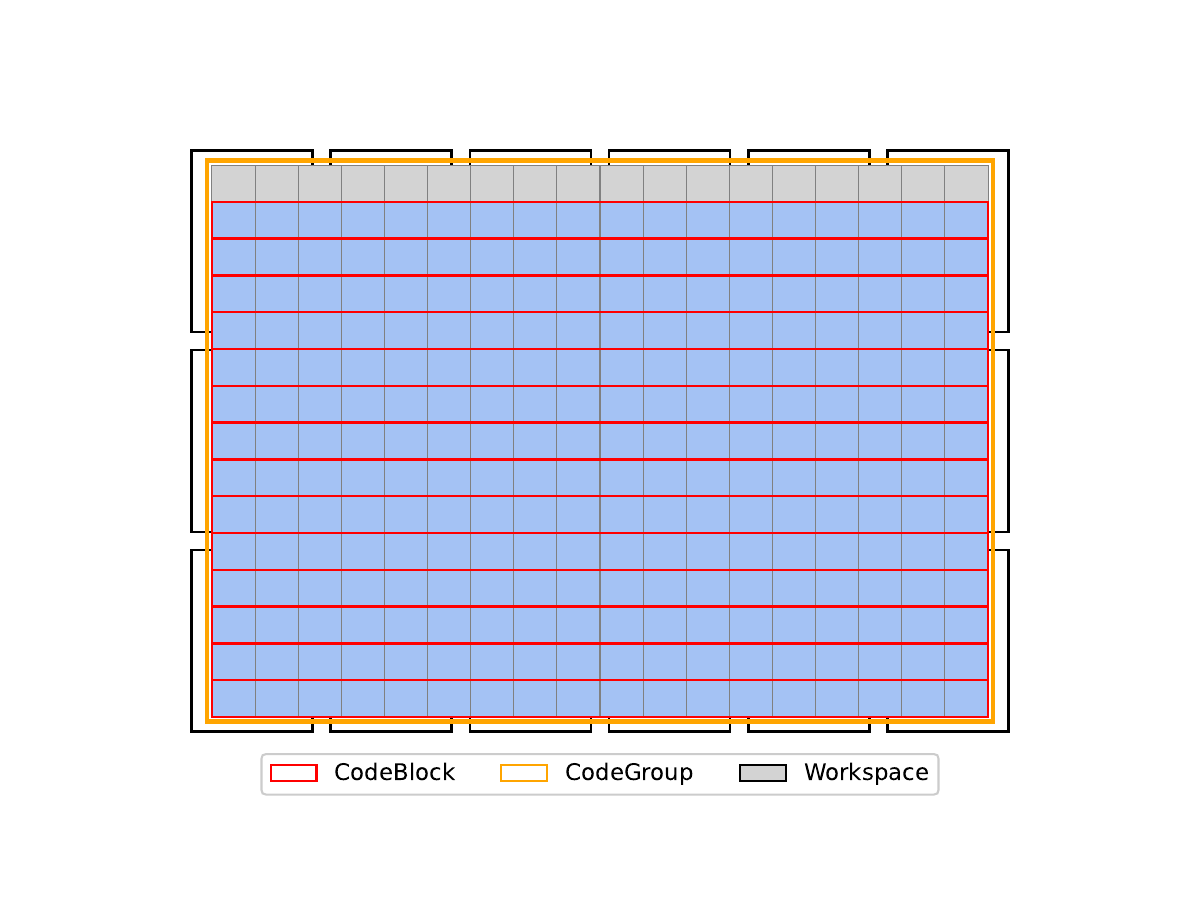}
        \hfill
        \includegraphics{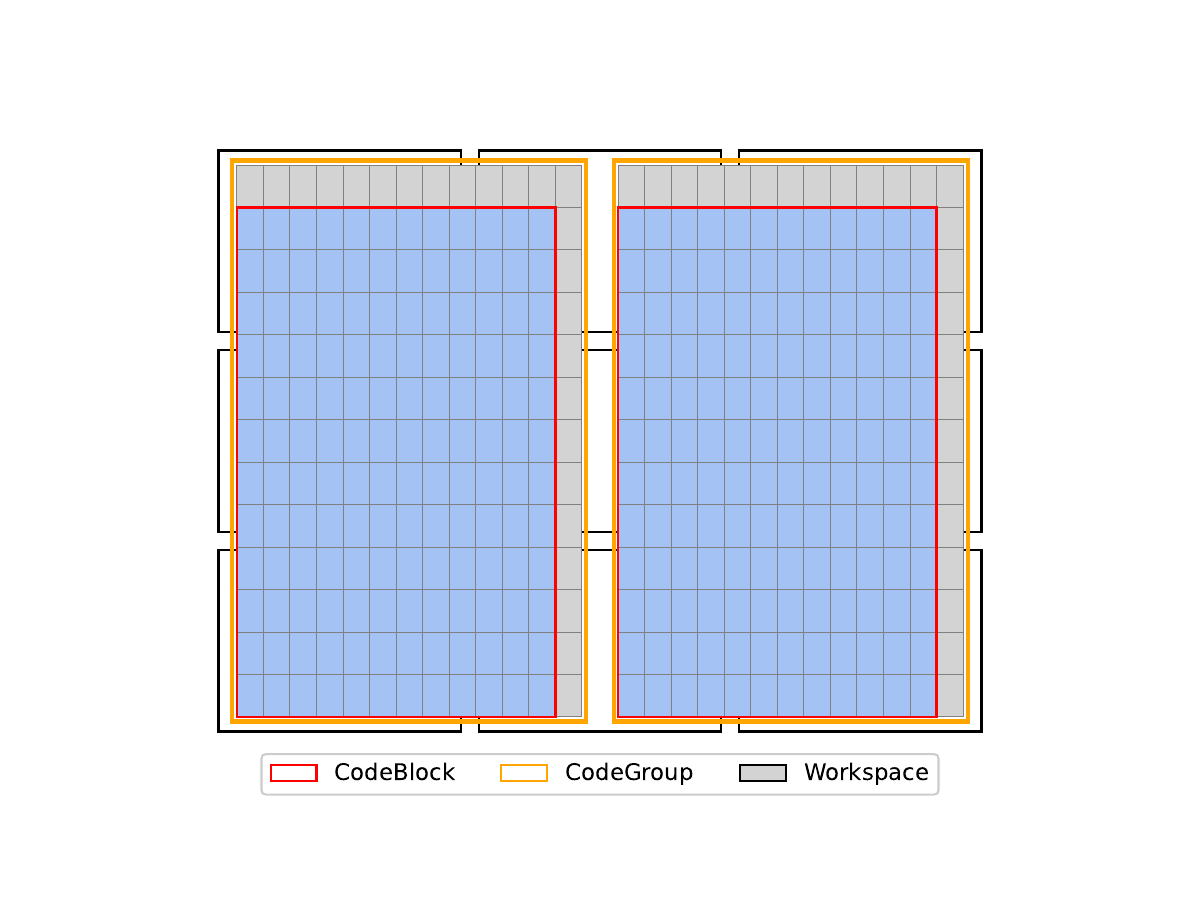}
    }\caption{Examples of \emph{CodeLayout}s for the memory zone in the DFTQC architecture. Left: a \emph{CodeLayout} using 1D yoked surface codes under chip size $h=3$, $w=2$, $d=31$. One \emph{CodeGroup} is placed on $3\times 6$ chips, and each \emph{CodeGroup} has one $[[18, 16, 2]]$ \emph{CodeBlock}. Right: a \emph{CodeLayout} using 2D yoked surface codes under chip size $h=3$, $w=4$, $d=31$. Two \emph{CodeGroup}s are placed on $3\times 3$ chips, and each \emph{CodeGroup} has one $[[144, 98, 4]]$ \emph{CodeBlock}.}\label{fig:CodeLayout}
\end{figure*}

In a realistic layout, a \emph{CodeGroup} may span multiple interfaces, both horizontal and vertical, with varying configurations for different \emph{CodeGroup}s. To manage this complexity, we introduce a new class, \emph{CodeLayout}, which organizes multiple \emph{CodeGroup}s into a 2D grid distributed across multiple chips which are also arranged in a 2D grid, as shown in \fref{fig:CodeLayout}. Then the memory zone is composed of repeated \emph{CodeLayout}s, as shown in the optimal layout for $(h=3,w=2,d=33)$ chips in \fref{fig:space_layout_demonstration}.

Note that, for the \emph{CodeLayout} of 1D yoked surface codes, because the parity check of each \emph{CodeBlock} may happen at different time, if the \emph{CodeBlock}s next to the horizontal interfaces are at different rows in their \emph{CodeGroup}s, there will be more rounds involving inter-chip coupling. The situation is similar for the 2D cases. 

For a \emph{CodeLayout}, there are inevitably some redundant qubits on some chips. These redundant qubits do not belong to any \emph{CodeGroup} and remain idle during the entire algorithm. To account for these qubits, we define the code rate of a \emph{CodeLayout} as the ratio of encoded logical qubits to the total number of physical qubits on all chips. This definition differs from the conventional definition of code rate, which only considers the qubits involved in encoding logical qubits. The \emph{CodeLayout}s with the maximum code rate under different target logical error rates for $(h=3,w=2,d=31)$ chips are shown in \fref{fig:footprint}, compared with the monolithic case.

\subsection{Fitting coefficients for lower physical error rates}\label{sec:fitting_coefficients_lower_p}

For lower physical error rates, we need to change the values of the fitting coefficients in \eref{eq:yoked_ler_with_interfaces}. As an approximation, we remain the values of $B_{n,k}$ and $B_{l,k}$ which we expect that will not change with the physical error rates dramatically. For $A_{n,k}$ and $A_{l,k}$, we note that the logical error rate should also scale as $(p/p_{th})^{(d+1)/2}$, where $p_{th}$ is the threshold of surface code under a given noise model, so $ A^{-d}\sim (p/p_{th})^{(d+1)/2}$ and $A\sim \sqrt{p_{th}/p}^{(d+1)/d} \sim \sqrt{p_{th}/p}$ . Thus we estimate the new values of $A_{n,k}$ and $A_{l,k}$ as $A_{n,k}'\sim \sqrt{p_b/p_b'}A_{n,k}$ and $A_{l,k}' \sim \sqrt{p_b/p_b'}A_{l,k}$.
Then we adopt the same procedure in \fref{fig:procedure} to search for the optimal configurations of yoked surface codes.

\bibliography{references}

@article{gidney_stim_2021,
	title = {Stim: a fast stabilizer circuit simulator},
	volume = {5},
	issn = {2521-327X},
	shorttitle = {Stim},
	url = {http://arxiv.org/abs/2103.02202},
	doi = {10.22331/q-2021-07-06-497},
	
	urldate = {2023-07-04},
	journal = {Quantum},
	author = {Gidney, Craig},
	month = jul,
	year = {2021},
	note = {arXiv:2103.02202 [quant-ph]},
	pages = {497},
}

@article{li_hierarchical_2016,
	title = {Hierarchical surface code for network quantum computing with modules of arbitrary size},
	volume = {94},
	copyright = {http://creativecommons.org/licenses/by/3.0/},
	issn = {2469-9926, 2469-9934},
	url = {https://link.aps.org/doi/10.1103/PhysRevA.94.042303},
	doi = {10.1103/PhysRevA.94.042303},
	
	number = {4},
	urldate = {2024-08-26},
	journal = {Phys. Rev. A},
	author = {Li, Ying and Benjamin, Simon C.},
	month = oct,
	year = {2016},
	pages = {042303},
}

@article{ramette_fault-tolerant_2024,
	title = {Fault-tolerant connection of error-corrected qubits with noisy links},
	volume = {10},
	issn = {2056-6387},
	url = {https://www.nature.com/articles/s41534-024-00855-4},
	doi = {10.1038/s41534-024-00855-4},
	
	number = {1},
	urldate = {2024-08-26},
	journal = {npj Quantum Inf},
	author = {Ramette, Joshua and Sinclair, Josiah and Breuckmann, Nikolas P. and Vuletić, Vladan},
	month = jun,
	year = {2024},
	pages = {58},
}

@article{niu_low-loss_2023,
	title = {Low-loss interconnects for modular superconducting quantum processors},
	volume = {6},
	issn = {2520-1131},
	url = {https://www.nature.com/articles/s41928-023-00925-z},
	doi = {10.1038/s41928-023-00925-z},
	number = {3},
	urldate = {2024-09-29},
	journal = {Nat Electron},
	author = {Niu, Jingjing and Zhang, Libo and Liu, Yang and Qiu, Jiawei and Huang, Wenhui and Huang, Jiaxiang and Jia, Hao and Liu, Jiawei and Tao, Ziyu and Wei, Weiwei and Zhou, Yuxuan and Zou, Wanjing and Chen, Yuanzhen and Deng, Xiaowei and Deng, Xiuhao and Hu, Changkang and Hu, Ling and Li, Jian and Tan, Dian and Xu, Yuan and Yan, Fei and Yan, Tongxing and Liu, Song and Zhong, Youpeng and Cleland, Andrew N. and Yu, Dapeng},
	month = feb,
	year = {2023},
	pages = {235--241},
}

@misc{gidney_magic_2024,
	title = {Magic state cultivation: growing {T} states as cheap as {CNOT} gates},
	shorttitle = {Magic state cultivation},
	url = {http://arxiv.org/abs/2409.17595},
	
	urldate = {2024-10-09},
	publisher = {arXiv},
	author = {Gidney, Craig and Shutty, Noah and Jones, Cody},
	month = sep,
	year = {2024},
	note = {arXiv:2409.17595 [quant-ph]},
}

@article{gidney_how_2021,
	title = {How to factor 2048 bit {RSA} integers in 8 hours using 20 million noisy qubits},
	volume = {5},
	issn = {2521-327X},
	url = {http://arxiv.org/abs/1905.09749},
	doi = {10.22331/q-2021-04-15-433},
	
	urldate = {2024-11-10},
	journal = {Quantum},
	author = {Gidney, Craig and Ekerå, Martin},
	month = apr,
	year = {2021},
	note = {arXiv:1905.09749 [quant-ph]},
	pages = {433},
}

@misc{heya_randomized_2025,
	title = {Randomized benchmarking of a high-fidelity remote {CNOT} gate over a meter-scale microwave interconnect},
	url = {http://arxiv.org/abs/2502.15034},
	doi = {10.48550/arXiv.2502.15034},
	urldate = {2025-02-25},
	publisher = {arXiv},
	author = {Heya, Kentaro and Phung, Timothy and Malekakhlagh, Moein and Steiner, Rachel and Turchetti, Marco and Shanks, William and Mamin, John and Lu, Wen-Sen and Kandel, Yadav Prasad and Sundaresan, Neereja and Orcutt, Jason},
	month = feb,
	year = {2025},
	note = {arXiv:2502.15034 [quant-ph]},
}

@misc{shalby_optimized_2025,
	title = {Optimized noise-resilient surface code teleportation interfaces},
	url = {http://arxiv.org/abs/2503.04968},
	doi = {10.48550/arXiv.2503.04968},
	urldate = {2025-03-11},
	publisher = {arXiv},
	author = {Shalby, Mohamed A. and Wang, Renyu and Sedov, Denis and Pryadko, Leonid P.},
	month = mar,
	year = {2025},
	note = {arXiv:2503.04968 [quant-ph]},
}

@misc{song_realization_2024,
	title = {Realization of high-fidelity perfect entangler between remote superconducting quantum processors},
	url = {http://arxiv.org/abs/2407.20338},
	doi = {10.48550/arXiv.2407.20338},
	
	urldate = {2025-03-20},
	publisher = {arXiv},
	author = {Song, Juan and Yang, Shuang and Liu, Pei and Zhang, Hui-Li and Xue, Guang-Ming and Mi, Zhen-Yu and Zhang, Wen-Gang and Yan, Fei and Jin, Yi-Rong and Yu, Hai-Feng},
	month = oct,
	year = {2024},
	note = {arXiv:2407.20338 [quant-ph]},
}

@article{gold_entanglement_2021,
	title = {Entanglement across separate silicon dies in a modular superconducting qubit device},
	volume = {7},
	issn = {2056-6387},
	url = {https://www.nature.com/articles/s41534-021-00484-1},
	doi = {10.1038/s41534-021-00484-1},
	
	number = {1},
	urldate = {2025-03-20},
	journal = {npj Quantum Inf},
	author = {Gold, Alysson and Paquette, J. P. and Stockklauser, Anna and Reagor, Matthew J. and Alam, M. Sohaib and Bestwick, Andrew and Didier, Nicolas and Nersisyan, Ani and Oruc, Feyza and Razavi, Armin and Scharmann, Ben and Sete, Eyob A. and Sur, Biswajit and Venturelli, Davide and Winkleblack, Cody James and Wudarski, Filip and Harburn, Mike and Rigetti, Chad},
	month = sep,
	year = {2021},
	pages = {142},
}

@article{qiu_deterministic_2025,
	title = {Deterministic quantum state and gate teleportation between distant superconducting chips},
	volume = {70},
	issn = {20959273},
	url = {https://linkinghub.elsevier.com/retrieve/pii/S209592732400879X},
	doi = {10.1016/j.scib.2024.11.047},
	
	number = {3},
	urldate = {2025-03-20},
	journal = {Science Bulletin},
	author = {Qiu, Jiawei and Liu, Yang and Hu, Ling and Wu, Yukai and Niu, Jingjing and Zhang, Libo and Huang, Wenhui and Chen, Yuanzhen and Li, Jian and Liu, Song and Zhong, Youpeng and Duan, Luming and Yu, Dapeng},
	month = feb,
	year = {2025},
	pages = {351--358},
}

@article{gidney_efficient_2019,
	title = {Efficient magic state factories with a catalyzed {\textbar}{CCZ}{\textgreater} to 2{\textbar}{T}{\textgreater} transformation},
	volume = {3},
	issn = {2521-327X},
	url = {http://arxiv.org/abs/1812.01238},
	doi = {10.22331/q-2019-04-30-135},
	
	urldate = {2025-03-31},
	journal = {Quantum},
	author = {Gidney, Craig and Fowler, Austin G.},
	month = apr,
	year = {2019},
	note = {arXiv:1812.01238 [quant-ph]},
	pages = {135},
}

@misc{gidney_how_2025,
	title = {How to factor 2048 bit {RSA} integers with less than a million noisy qubits},
	url = {http://arxiv.org/abs/2505.15917},
	doi = {10.48550/arXiv.2505.15917},
	urldate = {2025-05-25},
	publisher = {arXiv},
	author = {Gidney, Craig},
	month = may,
	year = {2025},
	note = {arXiv:2505.15917 [quant-ph]},
}

@misc{webster_pinnacle_2026,
	title = {The {Pinnacle} Architecture: Reducing the cost of breaking {RSA}-2048 to 100 000 physical qubits using quantum {LDPC} codes},
	url = {http://arxiv.org/abs/2602.11457},
	doi = {10.48550/arXiv.2602.11457},
	publisher = {arXiv},
	author = {Webster, Paul and others},
	month = feb,
	year = {2026},
	note = {arXiv:2602.11457},
}

@online{mundadaHeterogeneousArchitecturesEnable2026,
  title = {Heterogeneous Architectures Enable a 138x Reduction in Physical Qubit Requirements for Fault-Tolerant Quantum Computing under Detailed Accounting},
  author = {Mundada, Pranav S. and Khindanov, Aleksei and Wang, Yulun and Edmunds, Claire L. and Coote, Paul and Biercuk, Michael J. and Baum, Yuval and Hush, Michael},
  date = {2026-04-07},
  eprint = {2604.06319},
  eprinttype = {arXiv},
  eprintclass = {quant-ph},
  url = {http://arxiv.org/abs/2604.06319},
  urldate = {2026-04-09},
  pubstate = {prepublished}
}

@article{litinskiGameSurfaceCodes2019,
  title = {A {{Game}} of {{Surface Codes}}: {{Large-Scale Quantum Computing}} with {{Lattice Surgery}}},
  shorttitle = {A {{Game}} of {{Surface Codes}}},
  author = {Litinski, Daniel},
  date = {2019-03-05},
  journaltitle = {Quantum},
  shortjournal = {Quantum},
  volume = {3},
  pages = {128},
  issn = {2521-327X},
  url = {https://quantum-journal.org/papers/q-2019-03-05-128/},
  urldate = {2022-02-12},
  langid = {english}
}

@article{liMagicStatesFidelity2015,
  title = {A Magic State’s Fidelity Can Be Superior to the Operations That Created It},
  author = {Li, Ying},
  date = {2015-02-13},
  journaltitle = {New Journal of Physics},
  shortjournal = {New J. Phys.},
  volume = {17},
  number = {2},
  pages = {023037},
  issn = {1367-2630},
  url = {https://iopscience.iop.org/article/10.1088/1367-2630/17/2/023037},
  urldate = {2025-04-16},
  langid = {english}
}

@article{gidney_yoked_2025,
	title = {Yoked surface codes},
	volume = {16},
	issn = {2041-1723},
	url = {https://www.nature.com/articles/s41467-025-59714-1},
	doi = {10.1038/s41467-025-59714-1},
	
	number = {1},
	urldate = {2025-05-31},
	journal = {Nat Commun},
	author = {Gidney, Craig and Newman, Michael and Brooks, Peter and Jones, Cody},
	month = may,
	year = {2025},
	pages = {4498},
}

@misc{maurya_synchronization_2025,
	title = {Synchronization for {Fault}-{Tolerant} {Quantum} {Computers}},
	url = {http://arxiv.org/abs/2506.10258},
	doi = {10.1145/3695053.3730991},
	urldate = {2025-06-15},
	author = {Maurya, Satvik and Tannu, Swamit},
	month = jun,
	year = {2025},
	note = {arXiv:2506.10258 [quant-ph]},
}

@article{chamberland_universal_2022,
	title = {Universal quantum computing with twist-free and temporally encoded lattice surgery},
	volume = {3},
	issn = {2691-3399},
	url = {http://arxiv.org/abs/2109.02746},
	doi = {10.1103/PRXQuantum.3.010331},
	number = {1},
	urldate = {2025-09-14},
	journal = {PRX Quantum},
	author = {Chamberland, Christopher and Campbell, Earl T.},
	month = feb,
	year = {2022},
	note = {arXiv:2109.02746 [quant-ph]},
	pages = {010331},
}

@article{google_quantum_ai_and_collaborators_quantum_2025,
	title = {Quantum error correction below the surface code threshold},
	volume = {638},
	issn = {0028-0836, 1476-4687},
	url = {https://www.nature.com/articles/s41586-024-08449-y},
	doi = {10.1038/s41586-024-08449-y},
	
	number = {8052},
	urldate = {2025-09-24},
	journal = {Nature},
	author = {Google Quantum AI and Collaborators},
	month = feb,
	year = {2025},
	pages = {920--926},
}

@misc{haug_lattice_2025,
	title = {Lattice {Surgery} with {Bell} {Measurements}: {Modular} {Fault}-{Tolerant} {Quantum} {Computation} at {Low} {Entanglement} {Cost}},
	shorttitle = {Lattice {Surgery} with {Bell} {Measurements}},
	url = {http://arxiv.org/abs/2510.13541},
	doi = {10.48550/arXiv.2510.13541},
	urldate = {2025-10-16},
	publisher = {arXiv},
	author = {Haug, Trond Hjerpekjøn and Hillmann, Timo and Kockum, Anton Frisk and Laer, Raphael Van},
	month = oct,
	year = {2025},
	note = {arXiv:2510.13541 [quant-ph]},
}

@misc{singh_modular_2024,
	title = {Modular {Architectures} and {Entanglement} {Schemes} for {Error}-{Corrected} {Distributed} {Quantum} {Computation}},
	url = {http://arxiv.org/abs/2408.02837},
	doi = {10.48550/arXiv.2408.02837},
	urldate = {2025-10-16},
	publisher = {arXiv},
	author = {Singh, Siddhant and Gu, Fenglei and Bone, Sébastian de and Villaseñor, Eduardo and Elkouss, David and Borregaard, Johannes},
	month = aug,
	year = {2024},
	note = {arXiv:2408.02837 [quant-ph]},
}

@article{mcewen_relaxing_2023,
	title = {Relaxing {Hardware} {Requirements} for {Surface} {Code} {Circuits} using {Time}-dynamics},
	volume = {7},
	issn = {2521-327X},
	url = {http://arxiv.org/abs/2302.02192},
	doi = {10.22331/q-2023-11-07-1172},
	
	urldate = {2025-10-19},
	journal = {Quantum},
	author = {McEwen, Matt and Bacon, Dave and Gidney, Craig},
	month = nov,
	year = {2023},
	note = {arXiv:2302.02192 [quant-ph]},
	pages = {1172},
}

@misc{gidney_answer_2025,
	title = {Answer to "{Runtime} estimation in "{How} to factor 2048 bit {RSA} integers with less than a million noisy qubits""},
	url = {https://quantumcomputing.stackexchange.com/a/44571/31142},
	urldate = {2025-10-27},
	journal = {Quantum Computing Stack Exchange},
	author = {Gidney, Craig},
	month = sep,
	year = {2025},
}

@misc{oscarhiggott_oscarhiggottpymatching_2025,
	title = {oscarhiggott/{PyMatching}},
	copyright = {Apache-2.0},
	url = {https://github.com/oscarhiggott/PyMatching},
	urldate = {2025-10-27},
	author = {oscarhiggott},
	month = oct,
	year = {2025},
	note = {original-date: 2019-11-08T11:05:08Z},
}

@misc{jacinto_network_2025,
	title = {Network {Requirements} for {Distributed} {Quantum} {Computation}},
	url = {http://arxiv.org/abs/2504.08891},
	doi = {10.48550/arXiv.2504.08891},
	urldate = {2025-10-30},
	publisher = {arXiv},
	author = {Jacinto, Hugo and Gouzien, Elie and Sangouard, Nicolas},
	month = apr,
	year = {2025},
	note = {arXiv:2504.08891 [quant-ph]},
}

@misc{chandra_architectural_2025,
	title = {Architectural {Approaches} to {Fault}-{Tolerant} {Distributed} {Quantum} {Computing} and {Their} {Entanglement} {Overheads}},
	url = {http://arxiv.org/abs/2511.13657},
	doi = {10.48550/arXiv.2511.13657},
	urldate = {2026-01-13},
	publisher = {arXiv},
	author = {Chandra, Nitish Kumar and Kaur, Eneet and Seshadreesan, Kaushik P.},
	month = nov,
	year = {2025},
	note = {arXiv:2511.13657 [quant-ph]},
}

@misc{filippov_architecting_2025,
	title = {Architecting {Distributed} {Quantum} {Computers}: {Design} {Insights} from {Resource} {Estimation}},
	url = {https://arxiv.org/abs/2508.19160},
	doi = {10.48550/arXiv.2508.19160},
	publisher = {arXiv},
	author = {Filippov, Dmitry and Yang, Peter and Murali, Prakash},
	month = aug,
	year = {2025},
	note = {arXiv:2508.19160 [quant-ph]},
}

@misc{naito_network-based_2026,
	title = {Network-{Based} {Quantum} {Computing}: an efficient design framework for many-small-node distributed fault-tolerant quantum computing},
	shorttitle = {Network-{Based} {Quantum} {Computing}},
	url = {http://arxiv.org/abs/2601.09374},
	doi = {10.48550/arXiv.2601.09374},
	urldate = {2026-01-16},
	publisher = {arXiv},
	author = {Naito, Soshun and Suzuki, Yasunari and Tokunaga, Yuuki},
	month = jan,
	year = {2026},
	note = {arXiv:2601.09374 [quant-ph]},
}

@article{gidney_benchmarking_2022,
	title = {Benchmarking the {Planar} {Honeycomb} {Code}},
	volume = {6},
	issn = {2521-327X},
	url = {http://arxiv.org/abs/2202.11845},
	doi = {10.22331/q-2022-09-21-813},
	
	urldate = {2026-01-23},
	journal = {Quantum},
	author = {Gidney, Craig and Newman, Michael and McEwen, Matt},
	month = sep,
	year = {2022},
	note = {arXiv:2202.11845 [quant-ph]},
	pages = {813},
}

@misc{newville_lmfit_2025,
	title = {{LMFIT}: {Non}-{Linear} {Least}-{Squares} {Minimization} and {Curve}-{Fitting} for {Python}},
	url = {https://doi.org/10.5281/zenodo.16175987},
	doi = {10.5281/zenodo.16175987},
	publisher = {Zenodo},
	author = {Newville, Matthew and Otten, Renee and Nelson, Andrew and Stensitzki, Till and Ingargiola, Antonino and Allan, Daniel and Fox, Austin and Carter, Faustin and Rawlik, Michal},
	month = jul,
	year = {2025},
}

@online{mohseniHowBuildQuantum2024,
  title = {How to {{Build}} a {{Quantum Supercomputer}}: {{Scaling Challenges}} and {{Opportunities}}},
  shorttitle = {How to {{Build}} a {{Quantum Supercomputer}}},
  author = {Mohseni, Masoud and Scherer, Artur and Johnson, K. Grace and Wertheim, Oded and Otten, Matthew and Aadit, Navid Anjum and Bresniker, Kirk M. and Camsari, Kerem Y. and Chapman, Barbara and Chatterjee, Soumitra and Dagnew, Gebremedhin A. and Esposito, Aniello and Fahim, Farah and Fiorentino, Marco and Khalid, Abdullah and Kong, Xiangzhou and Kulchytskyy, Bohdan and Li, Ruoyu and Lott, P. Aaron and Markov, Igor L. and McDermott, Robert F. and Pedretti, Giacomo and Gajjar, Archit and Silva, Allyson and Sorebo, John and Spentzouris, Panagiotis and Steiner, Ziv and Torosov, Boyan and Venturelli, Davide and Visser, Robert J. and Webb, Zak and Zhan, Xin and Cohen, Yonatan and Ronagh, Pooya and Ho, Alan and Beausoleil, Raymond G. and Martinis, John M.},
  date = {2024-11-15},
  eprint = {2411.10406},
  eprinttype = {arXiv},
  eprintclass = {quant-ph},
  url = {http://arxiv.org/abs/2411.10406},
  urldate = {2024-11-28},
  langid = {english},
  pubstate = {prepublished},
  year = {2024},
}

@article{storz_loophole-free_2023,
	title = {Loophole-free {Bell} inequality violation with superconducting circuits},
	volume = {617},
	doi = {10.1038/s41586-023-05885-0},
	journal = {Nature},
	author = {Storz, Simon and Sch\"ar, Josua and Kulikov, Anatoly and Magnard, Paul and Kurpiers, Philipp and L\"utolf, Janis and Walter, Theo and Copetudo, Adrian and Reuer, Kevin and Akin, Abdulkadir and Motzoi, Jean-Claude and Eichler, Christopher and Wallraff, Andreas},
	year = {2023},
	pages = {265--270},
}

@article{magnard_microwave_2020,
	title = {Microwave Quantum Link between Superconducting Circuits Housed in Spatially Separated Cryogenic Systems},
	volume = {125},
	doi = {10.1103/PhysRevLett.125.260502},
	journal = {Phys. Rev. Lett.},
	author = {Magnard, Paul and Storz, Simon and Kurpiers, Philipp and Sch\"ar, Josua and Marxer, Fabian and L\"utolf, Janis and Walter, Theo and Besse, Jean-Claude and Gabureac, Mihai and Reuer, Kevin and Akin, Abdulkadir and Royer, Baptiste and Blais, Alexandre and Wallraff, Andreas},
	year = {2020},
	pages = {260502},
}

@article{conner_multi-chip_2023,
	title = {Modular Superconducting Qubit Architecture with a Multi-chip Tunable Coupler},
	url = {http://arxiv.org/abs/2308.09240},
	doi = {10.48550/arXiv.2308.09240},
	journal = {arXiv:2308.09240},
	author = {Conner, C. R. and Stehlik, J. and Zajac, D. M. and Brink, M. and Chow, J. M.},
	year = {2023},
}

@article{heya_multi-module_2025,
	title = {Performance Characterization of a Multi-Module Quantum Processor with Static Inter-Chip Couplers},
	url = {http://arxiv.org/abs/2503.12603},
	doi = {10.48550/arXiv.2503.12603},
	journal = {arXiv:2503.12603},
	author = {Heya, Kentaro and Kanazawa, Naoki and Suzuki, Yutaka and others},
	year = {2025},
}

@article{main_distributed_2025,
	title = {Distributed quantum computing across an optical network link},
	volume = {638},
	doi = {10.1038/s41586-024-08404-x},
	journal = {Nature},
	author = {Main, D. and Drmota, P. and Nadlinger, D. P. and Ainley, E. M. and Agrawal, A. and Nichol, B. C. and Srinivas, R. and Araneda, G. and Lucas, D. M.},
	year = {2025},
	pages = {383--388},
}

@article{krutyanskiy_entanglement_2023,
	title = {Entanglement of Trapped-Ion Qubits Separated by 230 Meters},
	volume = {130},
	doi = {10.1103/PhysRevLett.130.050803},
	journal = {Phys. Rev. Lett.},
	author = {Krutyanskiy, V. and Canteri, M. and Meraner, M. and Krcmarsky, V. and Lanyon, B. P.},
	year = {2023},
	pages = {050803},
}

@article{pompili_multinode_2021,
	title = {Realization of a multinode quantum network of remote solid-state qubits},
	volume = {372},
	doi = {10.1126/science.abg1919},
	journal = {Science},
	author = {Pompili, Matteo and Hermans, Sophie L. N. and Baber, Simon and Degen, Hans K. C. and Huang, Zhichao and Dirkse, Bart and Wehner, Stephanie and Hanson, Ronald},
	year = {2021},
	pages = {259--264},
}

@article{stolk_metropolitan_2024,
	title = {Metropolitan-scale heralded entanglement of solid-state qubits},
	volume = {10},
	doi = {10.1126/sciadv.adp6442},
	journal = {Sci. Adv.},
	author = {Stolk, Arian J. and others},
	year = {2024},
	pages = {eadp6442},
}

@article{knaut_entanglement_2024,
	title = {Entanglement of nanophotonic quantum memory nodes in a telecom network},
	volume = {629},
	doi = {10.1038/s41586-024-07252-z},
	journal = {Nature},
	author = {Knaut, C. M. and Suleymanzade, A. and Wei, Y.-C. and Assumpcao, D. R. and Stas, P.-J. and Huan, Y. Q. and Machielse, B. and Knall, E. N. and Sutula, M. and Barber, G. and Riedinger, R. and Landig, R. and Bhaskar, M. K. and Park, H. and Lon{\v c}ar, M. and Lukin, M. D.},
	year = {2024},
	pages = {573--578},
}

@article{warner_coherent_2025,
	title = {Coherent control of a superconducting qubit using light},
	doi = {10.1038/s41567-025-02812-0},
	journal = {Nat. Phys.},
	author = {Warner, Joshua M. and others},
	year = {2025},
}

@article{bravyiFutureQuantumComputing2022a,
  title = {The Future of Quantum Computing with Superconducting Qubits},
  author = {Bravyi, Sergey and Dial, Oliver and Gambetta, Jay M. and Gil, Darío and Nazario, Zaira},
  date = {2022-10-28},
  journaltitle = {Journal of Applied Physics},
  volume = {132},
  number = {16},
  pages = {160902},
  issn = {0021-8979, 1089-7550},
  url = {https://pubs.aip.org/jap/article/132/16/160902/2837574/The-future-of-quantum-computing-with},
  urldate = {2026-04-10},
  langid = {english}
}

@online{hugginsFLuidAllocationSurface2025,
  title = {The {{FLuid Allocation}} of {{Surface}} Code {{Qubits}} ({{FLASQ}}) Cost Model for Early Fault-Tolerant Quantum Algorithms},
  author = {Huggins, William J. and Khattar, Tanuj and Xu, Amanda and Harrigan, Matthew and Kang, Christopher and Low, Guang Hao and Fowler, Austin and Rubin, Nicholas C. and Babbush, Ryan},
  date = {2025-11-12},
  eprint = {2511.08508},
  eprinttype = {arXiv},
  eprintclass = {quant-ph},
  url = {http://arxiv.org/abs/2511.08508},
  urldate = {2025-11-12},
  pubstate = {prepublished}
}

@article{google_quantum_ai_suppressing_2023,
  title = {Suppressing quantum errors by scaling a surface code logical qubit},
  volume = {614},
  doi = {10.1038/s41586-022-05434-1},
  journal = {Nature},
  author = {{Google Quantum AI}},
  year = {2023},
  pages = {676--681},
}

@article{postler_demonstration_2022,
  title = {Demonstration of fault-tolerant universal quantum gate operations},
  volume = {605},
  doi = {10.1038/s41586-022-04721-1},
  journal = {Nature},
  author = {Postler, Lukas and Heu{\ss}en, Sascha and Pogorelov, Ivan and Rispler, Manuel and Feldker, Thomas and Meth, Michael and Marciniak, Christian D. and Stricker, Roman and Ringbauer, Martin and Blatt, Rainer and Schindler, Philipp and M{\"u}ller, Markus and Monz, Thomas},
  year = {2022},
  pages = {675--680},
}

@article{bluvstein_logical_2024,
  title = {Logical quantum processor based on reconfigurable atom arrays},
  volume = {626},
  doi = {10.1038/s41586-023-06927-3},
  journal = {Nature},
  author = {Bluvstein, Dolev and Evered, Simon J. and Geim, Alexandra A. and Li, Sophie H. and Zhou, Hengyun and Manovitz, Tom and Ebadi, Sepehr and Cain, Madelyn and Kalinowski, Marcin and Hangleiter, Dominik and Bonilla Ataides, Juan Pablo and Maskara, Nishad and Cong, Iris and Gao, Xun and Sales Rodriguez, Pedro and Karolyshyn, Tena and Semeghini, Giulia and Gullans, Michael J. and Greiner, Markus and Vuleti{\'c}, Vladan and Lukin, Mikhail D.},
  year = {2024},
  pages = {58--65},
}

@article{su_fault-tolerant_2021,
  title = {Fault-tolerant quantum simulations of chemistry in first quantization},
  volume = {2},
  doi = {10.1103/PRXQuantum.2.040332},
  journal = {PRX Quantum},
  author = {Su, Yuan and Berry, Dominic W. and Wiebe, Nathan and Rubin, Nicholas and Babbush, Ryan},
  year = {2021},
  pages = {040332},
}

@article{nickerson_topological_2013,
  title = {Topological quantum computing with a very noisy network and local error rates approaching one percent},
  volume = {4},
  doi = {10.1038/ncomms2773},
  journal = {Nat. Commun.},
  author = {Nickerson, Naomi H. and Li, Ying and Benjamin, Simon C.},
  year = {2013},
  pages = {1756},
}

@article{franke_rents_2019,
  title = {Rent's rule and extensibility in quantum computing},
  volume = {67},
  doi = {10.1016/j.micpro.2019.02.006},
  journal = {Microprocessors and Microsystems},
  author = {Franke, D. P. and Clarke, J. S. and Vandersypen, L. M. K. and Veldhorst, M.},
  year = {2019},
  pages = {1--7},
}

@online{yoderTourGrossModular2025,
  title = {Tour de Gross: {{A}} Modular Quantum Computer Based on Bivariate Bicycle Codes},
  shorttitle = {Tour de Gross},
  author = {Yoder, Theodore J. and Schoute, Eddie and Rall, Patrick and Pritchett, Emily and Gambetta, Jay M. and Cross, Andrew W. and Carroll, Malcolm and Beverland, Michael E.},
  date = {2025-06-03},
  eprint = {2506.03094},
  eprinttype = {arXiv},
  eprintclass = {quant-ph},
  url = {http://arxiv.org/abs/2506.03094},
  urldate = {2026-05-18},
  pubstate = {prepublished}
}

@article{bravyiHighthresholdLowoverheadFaulttolerant2024,
  title = {High-Threshold and Low-Overhead Fault-Tolerant Quantum Memory},
  author = {Bravyi, Sergey and Cross, Andrew W. and Gambetta, Jay M. and Maslov, Dmitri and Rall, Patrick and Yoder, Theodore J.},
  date = {2024-03-28},
  journaltitle = {Nature},
  shortjournal = {Nature},
  volume = {627},
  number = {8005},
  pages = {778--782},
  issn = {0028-0836, 1476-4687},
  url = {https://www.nature.com/articles/s41586-024-07107-7},
  urldate = {2024-05-24},
  langid = {english}
}
\end{document}